\documentclass[aps,prl,twocolumn,groupedaddress,citeautoscript,nopacs,10pt]{revtex4-2}
\usepackage[colorlinks=true]{hyperref}
 \usepackage{lipsum}
 \usepackage{graphicx}
 \usepackage{textcomp}
\usepackage[dvipsnames]{xcolor}
\usepackage{latexsym}
\usepackage{amsmath}
\usepackage{amsthm}
\usepackage{amssymb}
\usepackage{epstopdf}
\usepackage{enumitem}
\usepackage{soul}
\usepackage{setspace} % controllable line spacing
\usepackage{dcolumn} % Align table columns on decimal point
\usepackage{bm} % bold math
\usepackage{setspace} % controllable line spacing

\usepackage{slashed}
\usepackage{color}
\usepackage{xcolor}
\usepackage{youngtab}
\usepackage{tikz}
\usepackage{tikz-3dplot}
\usepackage{braket}
\usepackage{mhchem}
\usepackage{dsfont}

\usetikzlibrary{shapes,snakes,arrows,chains,matrix,positioning,scopes,calc}
\usetikzlibrary{decorations.markings}

\usepackage[tabbotcap]{subfigure}

\allowdisplaybreaks

\usepackage{lipsum}

\newlength{\diamondrulelength}
\newlength{\diamondrulethickness}
\begin{document}
\title{Interplay of competing bond-order and loop-current fluctuations as a possible mechanism for superconductivity in kagome metals}

\author{Asimpunya Mitra$^{1}$}
\email{asimpunya.mitra@mail.utoronto.ca}
\author{Daniel J. Schultz$^{2}$}
\author{Yong Baek Kim$^{1}$}
\email{yongbaek.kim@utoronto.ca}

\affiliation{$^1$Department of Physics, University of Toronto, Toronto, Ontario M5S 1A7, Canada}
\affiliation{$^2$Institute for Theoretical Condensed Matter Physics,
Karlsruhe Institute of Technology, 76131 Karlsruhe, Germany}
\date{\today}

\begin{abstract}
The pairing symmetry and underlying mechanism for superconducting state of \ce{AV3Sb5} (A=K, Rb, Cs) kagome metal has been a topic of intense investigation. In this work, we consider an 8-band minimal model, which includes V, and the two types of Sb, both within and above/below the kagome plane. This model captures the Fermi surface pocket with significant in-plane Sb contribution near the zone center, and also has the two types of van Hove singularities (VHS), one of which has a strong out of plane Sb weight. By including V-V and V-planar Sb nearest-neighbor Coulomb interactions, 
we obtain the susceptibilities for fluctuating bond-order and loop-current in both charge and spin channels, and examine the resulting superconducting instabilities. In particular, we find that the time-reversal odd (even) charge-loop-current (charge bond-order) fluctuations favor unconventional (conventional) pairing symmetry such as $s_{+-}$ and $d+id$ ($s_{++}$).
Recent experimental works have highlighted the presence of $s$-wave pairing with two distinct gaps, one isotropic and one anisotropic. We discuss how this scenario may be compatible with either $s_{++}$ or $s_{+-}$ pairing, with an isotropic gap on the pocket dominated by in-plane Sb, but a highly anisotropic gap on V-dominated bands.  
\end{abstract}
\maketitle

\textit{Introduction.--} The \ce{AV3Sb5} (A=K, Rb, Cs) kagome superconductors are an exciting class of compounds  which exhibit various symmetry-broken states \cite{Wilson2024,Jiang2022,Ortiz2019, Ortiz2020,Zhao2021,Jiang2021,Li_2022,Liege2024,Wang2021,Mielke_2022,Liu2022,Guguchia_2023,Asaba2024,Deng_2024, Wu2022, Ratcliff2021, Luo2023, Ge2024, Lou2022, Frachet2024, Xu2025, Xiao2023, Li2022, Zhang2023, Guguchia2023, Gupta2022, Ortiz2021, Wang2023,Wang2024, Deng2024, Liexp2022,Qian2021,Feng2025}, including charge orders and superconductivity (SC). While the presence of these broken symmetry states are well established, the nature of the mechanism driving the charge order and the SC states is still under active investigation \cite{Denner_2021,Wu_2021,Christensen2021,Lin2021,Park2021,Feng2021,Christensen2022,Romer_2022,Zhou2022,Ferrari2022,Lin_2022,Tazai_2022,Li_2023,Holbaek2023,Jiang2023,Yu2012,Dong2023,Wu_2023,Wagner2023,Ritz2023,Tazai2023,Wu2023,Li_2024,Fu_2024,Dai2024,Schwemmer2024,Lin2024,Sim2024,Ojajarvi2024,Li_2025,Tian2025,Schultz_2025,Hoelbaek_2025,Han2025,tazai2025,Luscher2025,wang2025,Das2025}.

A plethora of experiments support the formation of charge bond-order \cite{Zhao2021,Jiang2021,Li2022_2,Shumiya2021,Chen2022,Li2021,Li_2022}, with charge density modulations on the vanadium (V) bonds. Further, time-reversal (TR) symmetry breaking has been reported in the charge-ordered \cite{Jiang2021,Wang2021,Shumiya2021,Wu_2022,Mielke_2022,Khasanov2022,Graham2024,Guguchia_2023,Xu2022,Asaba2024} state, despite the absence of local-moment spin order \cite{Kenney_2021,Ortiz2019}, thereby suggesting loop-current order as a possible origin of the TR-breaking. However, a number of other studies \cite{Li2022_2,Graham2024,Guo2024,Liege2024} have been unable to detect a stable long-range loop current order, leaving its status contested. 
These observations, however, do not preclude the existence of fluctuating charge loop-currents, or other fluctuating modes in the spin channel. Further, applying pressure \cite{Tsirlin2023,Stier2024,Yu2021,Chen2021,Wang2024} or doping \cite{Oey2022,Deng_2024,Kautzsch2023}  suppresses and eventually destroys the charge order. Hence, in this disordered state, various fluctuating bond-orders (BO) and loop-currents (LC) all coexist, and the effect of their competing fluctuations on SC in the presence of a realistic band structure is yet to be explored.

%Further, applying pressure \cite{Tsirlin2023,Stier2024,Yu2021,Chen2021,Wang2024} or doping \cite{Oey2022,Deng_2024,Kautzsch2023} drives these system into a disordered regime, where even the charge-bond order becomes a fluctuating mode. Consequently, in the disordered state, various fluctuating bond-orders (BO) and loop-currents (LC) may coexist, however the impact of these coexisting fluctuations on the SC has not been well explored.

\begin{figure}
    \centering
    \includegraphics[scale=0.8]{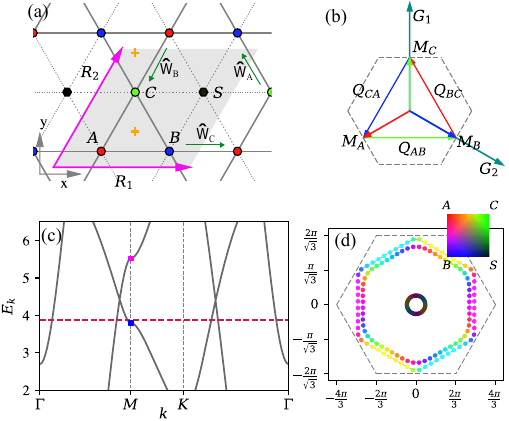}
  \caption{(a) The kagome plane, with V at $A,B,C$, in-plane-Sb at $S$, out-of-plane-Sb at the + positions within the grey unit cell. The green arrows $\Hat{\text{w}}_{A/B/C}$ show the different bond directions. (b) In the Brillouin zone reciprocal lattice vectors $Q_{AB}\equiv M_{C}$ and so on (same colour). (c) The electronic band structure of the 8-band model with the P- and M-type VHSs (blue and pink dots). (d) The 3-sheet Fermi surface for $\mu=3.88$ (dashed red line), the color scale denotes the weights of V-orbitals at sub-lattices $A,B,C$, and the in-plane-Sb orbital at $S$. We define, $\mathbf{x}_C=\mathbf{R}_1$, $\mathbf{x}_A=\mathbf{R}_2-\mathbf{R}_1$, $\mathbf{x}_B=-\mathbf{R}_2$.} 
\label{fig:1_main_text}
\end{figure}

In this letter, we present a microscopic study of the pairing interaction in the kagome metals mediated by fluctuations of BO and LCs in the charge and spin channels, all of which originate from nearest-neighbor Coulomb interactions. Inspired by \cite{Schultz_2025}, we include BO and LC order parameters between V and in-plane Sb (Sb-ip) sites as well as those between nearest neighbor V sites. First, we construct an effective microscopic model that captures the van-Hove singularities (VHS)
and a Fermi surface (FS) near the zone center
with appropriate orbital contents.
Using this model, we compute the RPA-corrected susceptibilities in the disordered state and identify various fluctuating BO \& LC channels close to the P-type VHS which are strongly enhanced near the $M$ point in the Brillouin zone. Finally, by tuning the relative strengths of the various fluctuating channels incorporated in the RPA-corrected Cooper vertex, we show that the different fluctuations compete with each other to mediate distinct types of SC. 
The properties of the resulting superconducting states are discussed in light of recent experiments.

\textit{Electronic structure.--}  
The electronic band structure of the \ce{AV3Sb5} kagome metals features multiple VHSs at the $M$ points close to the Fermi energy \cite{Wilson2024,Jiang2022}. The VHS mostly come from the V $3$-d orbitals \cite{Ortiz2020,Ortiz2019,Nakayama2021,Tan2021,LaBollita2021,Kang_2022,Luo2022,Hu2022,Jeong_2022,Li_2023,Zeng2025}, but a strong hybridization between the out-of-plane Sb (Sb--op) and V is responsible for forming the upper M-type VHS \cite{Jeong_2022,Li_2023,Tsirlin2022,Ritz_2023,Tsirlin2025}. Close to van-Hove fillings, the approximate nesting wave vector $\mathbf{Q}_{ab}$ (Fig.~\ref{fig:1_main_text} (b)) enhances various susceptibilities at the charge-ordering wave vector $M_{c}$, which has been proposed to be the driver of the charge BO formation \cite{Dong2023,Kiesel_2013,Wang2013}. It has been shown that the putative LC states may primarily emerge from the mirror-odd $\sigma_h=-1$ orbitals \cite{Li_2024,Li_2023} and distinct mirror symmetries of the wavefunction at the P- and M-type VHSs help in stabilizing the LC ordered state \cite{Li_2024,Li_2023}. In the disordered state, this would likely enhance the LC fluctuations. In this study, we construct an 8-band tight-binding model, with orbitals in the $\sigma_h=-1$ sub-sector. This model, based on the DFT band structure from Ref. \cite{Li_2023}, includes one effective V-$3d$ orbital per V site, Sb-$5p_z$ from the Sb-ip, and four linear combinations of the Sb-op $5p$ orbitals (see Supplement \cite{supplement} for details of the model). The band structure of this model in Fig.~\ref{fig:1_main_text} (c), features the P- and M-type VHSs with the correct mirror symmetries as in \cite{Li_2024}. It also has a circular Fermi $\Gamma$-pocket which is primarily composed of the Sb-ip $p_z$ orbitals (Fig.~\ref{fig:1_main_text} (d)). Removal of this $\Gamma$-pocket through a Lifshitz transition at 7.5 GPa coincides with the disappearance of superconductivity in low pressure regime of \ce{CsV3Sb5} \cite{Chen_2021,Zhang2021,Oey2022,Wilson2024}, hinting towards its 
importance in the low-pressure SC state. Also, quasi-particle interference spectroscopy has observed a large gap on this Sb-ip dominated $\Gamma$-pocket in \ce{KV3Sb5} \cite{Deng2024}. This crucial $\Gamma$-pocket, and the two VHSs with the correct mirror symmetries, are built into our 8-band model.

\textit{Interaction Hamiltonian.-} 
We choose the chemical potential $\mu$ to be close to a P-type VHS (see Figs.~\ref{fig:1_main_text}(c, d)) so that effects of the on-site Hubbard-U are suppressed due to sub-lattice interference \cite{Kiesel_2012,Kiesel_2013}. We therefore focus on two kinds of nearest-neighbor (NN) Coulomb repulsions: $\mathcal{V}_{\text{vv}}$ between the V (v) sites and $\mathcal{V}_{\text{sv}}$ between the V and Sb-ip (s) sites (for details, see %notation explained in 
Supplement \cite{supplement}),
\begin{gather}
H_{\text{int}}
=\mathcal{V}_{\text{vv}} \sum_{\mathcal{\mathbf{r}}_j,(a,b,c)} \left(n^{\text{v}}_{\mathcal{\mathbf{r}}_j,a} n^{\text{v}}_{\mathcal{\mathbf{r}}_j,b}+n^{\text{v}}_{\mathcal{\mathbf{r}}_j,a} n^{\text{v}}_{\mathcal{\mathbf{r}}_j-\mathbf{x}_{c},b}\right)\nonumber\\
 +  \mathcal{V}_{\text{sv}} \sum_{\mathcal{\mathbf{r}}_j,(a,b,c)} n^{\text{s}}_{\mathcal{\mathbf{r}}_j+\mathbf{\Tilde{x}}_{a}} \left(n^{\text{v}}_{\mathcal{\mathbf{r}}_j,b}+ n^{\text{v}}_{\mathcal{\mathbf{r}}_j,c} \right)  \label{eq:NN_Coulomb_interacting_ham},
\end{gather}
where $(a,b,c)$ denotes cyclic permutations of the $(A,B,C)$ and 
$(\mathbf{{\Tilde{x}}}_A, \mathbf{{\Tilde{x}}}_B, \mathbf{{\Tilde{x}}}_C) = 
(0, -\mathbf{x}_C, \mathbf{x}_B)$. $\mathbf{x}_A, \mathbf{x}_B, \mathbf{x}_C$ are defined in Fig.\ref{fig:1_main_text} caption.
In the unit cell $\mathcal{\mathbf{r}}_j$, $n^{\text{v}}_{\mathcal{\mathbf{r}}_j,a}$ is the local density of V at sub-lattice $a$, and $n^{\text{s}}_{\mathcal{\mathbf{r}}_j}$ is the local density of the Sb-ip. 
%Since in our 8-band model, both V and Sb-ip each contribute one orbital per sub-lattice site in the unit cell, the sub-lattice index can be regarded as the orbital index. 
We can reorganize Eq.~\eqref{eq:NN_Coulomb_interacting_ham} as an interaction between different types of bond-order and loop-current channels. In this decomposition, in addition to the conventionally studied \textit{charge} bond-order (cBO) and \textit{charge} loop-current (cLC) channels \cite{Denner_2021,Li_2023,Li_2024,Dong2023}, \textit{spin} bond-order (sBO) and \textit{spin} loop-current (sLC) channels are also present. We define four distinct types of order parameters: \textit{charge bond-orders:} $\mathcal{\hat{B}}^{\text{c,vv}}_{ \mathbf{q}}$, $\mathcal{\hat{B}}^{\text{c,sv}}_{ \mathbf{q}}$; \textit{charge loop-currents}: $\mathcal{\hat{J}}^{\text{c,vv}}_{ \mathbf{q}}$, $\mathcal{\hat{J}}^{\text{c,sv}}_{ \mathbf{q}}$; \textit{spin bond-orders}: $\mathcal{\hat{B}^{\text{s,vv}}_{ \mathbf{q}}}$, $\mathcal{\hat{B}^{\text{s,sv}}_{ \mathbf{q}}}$; and \textit{spin loop-currents:} $\mathcal{\hat{J}^{\text{s,vv}}_{ \mathbf{q}}}$, $\mathcal{\hat{J}^{\text{s,sv}}_{ \mathbf{q}}}$, with each type allowed between either V-V (vv) or V-Sb-ip (sv). Each order parameter is composed of three components, \textit{e.g.} $\mathcal{\hat{B}}^{\text{c,vv}}_{ \mathbf{q}}=\left(\mathcal{\hat{B}}^{\text{c,vv}}_{\mathbf{q},A},\mathcal{\hat{B}}^{\text{c,vv}}_{\mathbf{q},B},\mathcal{\hat{B}}^{\text{c,vv}}_{\mathbf{q},C}\right)$, along the three bond directions ($\hat{\text{w}}$) shown in Fig.~\ref{fig:1_main_text}(a). For each component, condensation or ordering may occur for a different momentum $\mathbf{q}=\mathbf{q}_A, \mathbf{q}_B, \mathbf{q}_C$. 
The V-V BO \& LC order-parameter along the $\hat{\text{w}}_C$ direction in Fig.~\ref{fig:1_main_text} (a) is,
\begin{align}
\hspace{-0.08cm}\mathcal{\hat{B}}^{\texttt{i}\text{,vv}}_{ \mathbf{q},C}\hspace{-0.1cm}=\hspace{-0.05cm}\sum_{\mathbf{k}\sigma\sigma'}\left[X^{A\sigma,B\sigma'}_{\mathbf{k},\mathbf{k+q}}f^{\text{vv}}_{C,\mathbf{k+q}}+X^{B\sigma,A\sigma'}_{\mathbf{k},\mathbf{k+q}}f^{\text{vv}}_{C,\mathbf{-k}} \right]\frac{\Gamma^\texttt{i}_{\sigma\sigma'}}{2}\label{eq:BO_main_text}\\
\hspace{-0.08cm}\mathcal{\hat{J}}^{\texttt{i}\text{,vv}}_{\mathbf{q},C}\hspace{-0.1cm}=\hspace{-0.1cm}i\hspace{-0.05cm}\sum_{\mathbf{k}\sigma\sigma'} \left[ X^{B\sigma,A\sigma'}_{\mathbf{k},\mathbf{k+q}}f^{\text{vv}}_{C,\mathbf{-k}} -X^{A\sigma,B\sigma'}_{\mathbf{k},\mathbf{k+q}}f^{\text{vv}}_{C,\mathbf{k+q}}\right]\frac{\Gamma^\texttt{i}_{\sigma\sigma'}}{2}\label{eq:LC_main_text}
\end{align}
where, $X^{A\sigma,B\sigma'}_{\mathbf{\mathbf{p},q}}=c^{\dagger}_{\mathbf{p},{\cal O}(A),\sigma}c_{\mathbf{q},
{\cal O}(B),\sigma'}$ with ${\cal{O}}(A),{\cal{O}}(B)$
referring to the V orbitals at the sublattice sites $A, B$
(one effective $d$-orbital per V sub-lattice site) and spin indices $\sigma,\sigma'$. $f^{\text{vv}}_{C,\mathbf{k}}=1-e^{-i\mathbf{k}\cdot \mathbf{x}_C}$ is a form factor associated with the order parameter along the $\hat{\text{w}}_C$ direction (see Fig.~\ref{fig:1_main_text} (a)). In Eqs.~\eqref{eq:BO_main_text},~\eqref{eq:LC_main_text}, $\texttt{i}\in \{c,s\}$ specify the charge ($\mathds{1}$) and spin ($\Vec{\boldsymbol{\sigma}}$) channels respectively, i.e. $(\Gamma^{\text{c}}, \Gamma^{\text{s}}) = (\mathds{1}, \Vec{\boldsymbol{\sigma}})$. The order parameters along the other bond directions, or in V-Sb-ip channels (sv) can be defined similarly (see Supplement \cite{supplement}). The bare interaction Hamiltonian can be rewritten as 
\begin{align}
\hspace{-0.1cm}H_{\text{int}}=-&\frac{\mathcal{V}_{\text{vv}}}{2N}\sum_{\mathbf{q},\texttt{i}\in \{c,s\}} \left(\mathcal{\hat{B}}^{\texttt{i},\text{vv}}_{\mathbf{q}}\cdot\mathcal{\hat{B}}^{\texttt{i},\text{vv}}_{\mathbf{-q}}+
\mathcal{\hat{J}}^{\texttt{i},\text{vv}}_{\mathbf{q}}\cdot\mathcal{\hat{J}}^{\texttt{i},\text{vv}}_{\mathbf{-q}}
\right)\nonumber\\
-&\frac{\mathcal{V}_{\text{sv}}}{2N}\sum_{\mathbf{q},\texttt{i}\in \{c,s\}}\left(
\mathcal{\hat{B}}^{\texttt{i},\text{sv}}_{\mathbf{q}}\cdot\mathcal{\hat{B}}^{\texttt{i},\text{sv}}_{\mathbf{-q}}+
\mathcal{\hat{J}}^{\texttt{i},\text{sv}}_{\mathbf{q}}\cdot\mathcal{\hat{J}}^{\texttt{i},\text{sv}}_{\mathbf{-q}}
\right).\label{eq:final_interaction_ham_split}
\end{align} 
In the subsequent sections, we obtain the RPA-corrected effective interaction, which incorporates FS features into the vertex.

\begin{figure*}
    \centering
    \includegraphics[scale=0.9]{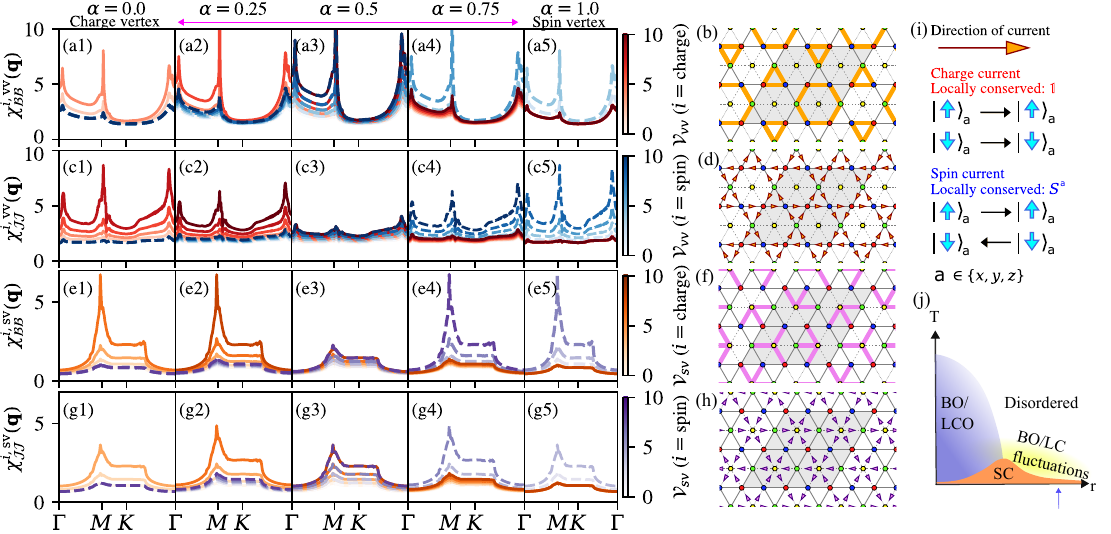}
    \caption{The RPA corrected susceptibilities (at $\mu=3.88$) for: bond-orders $\chi_{\mathcal{BB}}(\mathbf{q})$ ((a) between V-V, (e) between V-Sbip), and loop-currents $\chi_{\mathcal{JJ}}(\mathbf{q})$ ((c) between V-V, (g) between V-Sbip). Columns 1-5 show the susceptibilities in the charge and spin channels (different color scales) with increasing interaction strengths (shades of colors) as the relative strength of the charge and spin vertex is tuned by $\alpha$. Only the $\chi$’s in the disordered state are shown, some larger interaction strengths (or darker color shades) are omitted as those channels would condense. In (a,c) $\chi_{\mathcal{BB}}^{\texttt{i},\text{vv}}$ exceeds $\chi_{\mathcal{JJ}}^{\texttt{i},\text{vv}}$, and in (e,g) $\chi_{\mathcal{JJ}}^{\texttt{i},\text{sv}}$ exceeds $\chi_{\mathcal{BB}}^{\texttt{i},\text{sv}}$, when compared at the same interaction strengths (having same color shade).  The ordered state resulting from condensation at the three $\mathbf{q}=M_{c}$ (or $3\mathbf{Q}$ states)  for each BO \& LC channel is shown on the right in (b,d,f,h). (i) A charge (spin) current corresponds to a locally conserved density $\mathds{1}$ (spin $S^{\textsf{a}}$), with spin up/down  (in $\textsf{a}$-basis) flowing in the same (opposite) directions. (j) Fluctuations computed in the disordered state (blue arrow) where r is pressure/doping axis.}
    \label{fig:2_main_text}
\end{figure*}

\textit{Susceptibilities.--} We compute the static ($\omega=0$) RPA-corrected susceptibilities for bond-orders $\chi^{\texttt{i}}_{\mathcal{BB}}(\mathbf{q})=\left \langle \mathcal{\hat{B}}_{\mathbf{q}}^{\texttt{i}} \mathcal{\hat{B}}_{\mathbf{-q}}^{\texttt{i}} \right\rangle$ and loop-currents $\chi^{\texttt{i}}_{\mathcal{JJ}}(\mathbf{q})=\left \langle \mathcal{\hat{J}}_{\mathbf{q}}^{\texttt{i}} \mathcal{\hat{J}}_{\mathbf{-q}}^{\texttt{i}} \right\rangle$, for both the V-V and V-Sb-ip channels in the disordered state (see Fig.~\ref{fig:2_main_text} (j)). To gain more insight, we have parameterized the charge-spin sector of the vertex by, $V_{\text{charge-spin}}(\alpha)=(1-\alpha)\Gamma^{\text{c}}\Gamma^{\text{c}}+\alpha\Gamma^{\text{s}}\cdot \Gamma^{\text{s}}$, where $\alpha$ tunes the relative strengths of the charge and spin channels. While the NN-Coulomb interaction fixes $\alpha=\frac{1}{2}$, electron–phonon couplings \cite{Wu2024,You2025,Zhong2023,Yang2024} can potentially renormalize $\alpha$. The orbital-sector of the vertex remains unchanged.

\begin{figure*}
\centering\includegraphics[scale=0.9]{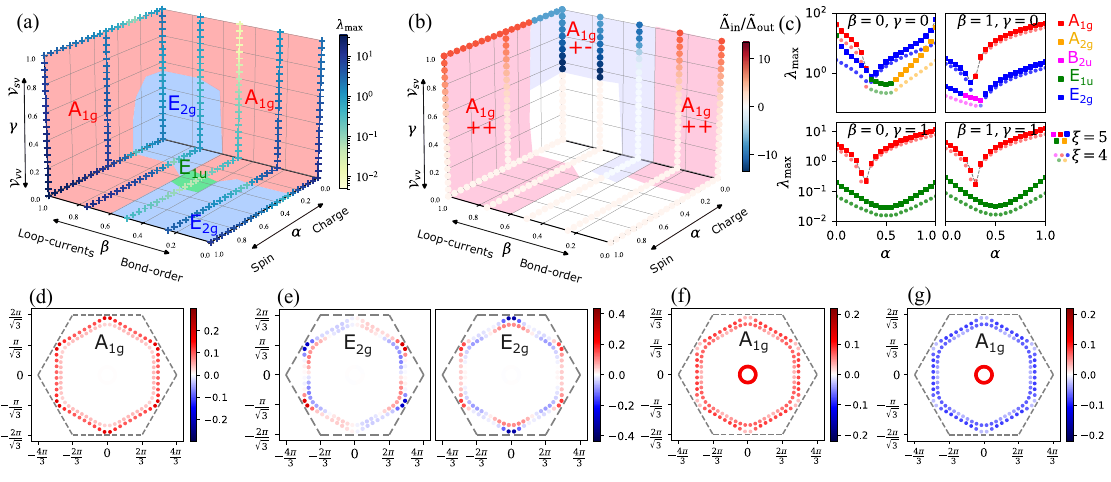}
    \caption{SC phase diagram: (a) as a function of the strength of the different fluctuating channels parameterized by $\alpha$ (charge $\leftrightarrow$ spin), $\beta$ (bond-order $\leftrightarrow$ loop-current), $\gamma$ (V-V $\leftrightarrow$ V-Sbip) for an interaction scale $\xi=4$. (b) Sub-classification of the $A_{1g}$ phase into the $s_{++}$ type and $s_{+-}$ type based on the value of $\Tilde{\Delta}_{\text{in}}/\Tilde{\Delta}_{\text{out}}$ (defined in text). 
    (c) The leading and sub-leading  SC instabilities for two different interaction scales $\xi=4$ (circles), $\xi=5$ (squares). The gap structure of the leading SC instability for fluctuations of: (d) V-V cBO ($\alpha=\beta=\gamma=0$), (e) V-V cLC ($\alpha=\gamma=0, \beta=1$), (f) V-Sbip cBO ($\alpha=\beta=0, \gamma=1$), (g) V-Sbip cLC ($\alpha=0, \beta=\gamma=1$).
    }\label{fig:Fig_3_main_text}
\end{figure*}

In Figs.~\ref{fig:2_main_text} (a,c,e,g), we show the BO \& LC susceptibilities, i.e. $\chi_{\mathcal{BB}}$ and $\chi_{\mathcal{JJ}}$ respectively in both V-V (vv) and the V-Sbip (sv) channels for $\mu=3.88$ near the P-type VHS. As seen from the columns in Fig.~\ref{fig:2_main_text}, tuning the interaction vertex $\alpha$ between the charge and spin channels respectively enhances the charge ($\mathtt{i}=c$)  and spin ($\mathtt{i}=s$) susceptibilities (shown with different color scales in each panel in Fig.~\ref{fig:2_main_text}). Additionally, as a general trend we see that both the BO \& LC susceptibilities in all the channels become most strongly peaked close to all the three $M_c$-points (see Fig.~\ref{fig:1_main_text} (b), where $c\in \{A,B,C\}$) with increasing interaction strengths. Further, the strengths of the fluctuations are roughly comparable as evident from the peak heights. For V-V charge orders (see Fig.~\ref{fig:2_main_text} (a1)), this $M_{c}$-ordering tendency is in agreement with experiments \cite{Wang2021,Jiang2021,Li_2022} which detect the formation of a $2\times 2$ charge order on the V kagome net. Further, in Figs.~\ref{fig:1_main_text} (a,c) we find that the V-V BO susceptibilities $\chi^{\texttt{i},\text{vv}}_{\mathcal{BB}}(\mathbf{q})$ are larger than the V-V LC susceptibility $\chi^{\texttt{i},\text{vv}}_{\mathcal{JJ}}(\mathbf{q})$, in agreement with previous studies which find BO phase close to the P-type VHS \cite{Li_2024,Dong2023,Ferrari2022}. We also find that moving $\mu$ closer to the M-type VHS results in an increase in the ratio of the LC to BO susceptibility or $\chi^{\texttt{i},\text{vv}}_{\mathcal{JJ}}(\mathbf{q})$/$\chi^{\texttt{i},\text{vv}}_{\mathcal{BB}}(\mathbf{q})$ at $\mathbf{q}=M_{c}$ which is caused by enhanced LC fluctuations. This is expected as  Refs. \cite{Li_2023,Li_2024} found the mean-field LC-order to be stabilized close to the M-type VHS. However, on the V-Sb-ip bonds, the $M_{c}$-peaked V-Sb-ip LC susceptibility $\chi_{\mathcal{JJ}}^{\texttt{i},\text{sv}}(\mathbf{q})$ is stronger than the V-Sb-ip BO susceptibility $\chi_{\mathcal{BB}}^{\texttt{i},\text{sv}}(\mathbf{q})$ for the same interaction strength (see Figs.~\ref{fig:2_main_text} (e,g)). We show in Figs. ~\ref{fig:2_main_text} (b, d, f, h), the pattern of the different long-range ordered states if they were allowed to condense at all the three $\mathbf{q}=M_{c}$, resulting in a $3\mathbf{Q}$ ordered state (i.e. for the three components: $\mathbf{q}_A=M_A$, $\mathbf{q}_B=M_B$, and $\mathbf{q}_C=M_C$ respectively). The cBOs have modulations of bond charge densities, and the sBOs have alternating bonds of increased spin-up and spin-down occupation \cite{Kiesel_2013,Profe2024}. The LCs (in Figs.~\ref{fig:2_main_text} (d, h)) preserve Kirchhoff's law and also the Bloch's global-current constraint \cite{Palle_2024,Schultz_2025,Palle_thesis}. Additionally, the cLCs conserve the local density $n_{\mathcal{\mathbf{r}}_j}$, and the sLCs conserve the local spin $S^a_{\mathcal{\mathbf{r}}_j}$ (see Fig.~\ref{fig:2_main_text} (i)). In Fig.~\ref{fig:2_main_text} (a,c,e,g) , the spin and charge susceptibilities are equal for $\alpha=0.5$. However, inclusion of beyond RPA-bubble diagrams are expected to comparatively strengthen the charge channels \cite{Romer_2022}. This would be equivalent to reducing $\alpha$, while leaving the other results unchanged.

Since each kind of fluctuation can act as a pairing glue to potentially distinct kind of SC \cite{Palle_2024,Schultz_2025}, 
we henceforth study the phase diagram as we tune between different fluctuating channels, by adjusting their relative strengths using $\{\alpha,\beta, \gamma\}$. As defined earlier, $\alpha$ tunes between the charge and spin channels. The relative strength of the bond-order and loop-current channel is tuned by $\beta$: $V_{\text{BO+LC}}(\beta)=(1-\beta)V_{\text{BO}}+\beta V_{\text{LC}}$. We do not consider mixing among the BO and LC channels in the RPA susceptibility \cite{supplement}. $\gamma$ tunes between the V-V and V-Sb-ip interactions: $\mathcal{V}_{\text{vv}}=(1-\gamma)\xi$ and $\mathcal{V}_{\text{sv}}=\gamma\xi$ in Eq.~\eqref{eq:final_interaction_ham_split}, here $\xi$ fixes the overall interaction scale.

\textit{SC phase diagram.--} 
The leading SC instability is found by solving a linearized gap equation using the RPA-corrected Cooper vertex incorporating the different fluctuating modes. Since we only focus on the competition between different channels, we set the overall interaction scale $\xi=4$. See End Matter for more details. The various SC instabilities at $\mu=3.88$, as a function of the three parameters $\{\alpha, \beta, \gamma\}$ are shown in Fig.~\ref{fig:Fig_3_main_text} (a). Most of the parameter space is dominated by singlet $A_{1g}$ ($s$-wave) and $E_{2g}$ ($d$-wave) channels. This is in agreement with experiments which report evidence of singlet pairing \cite{Mu_2021}, although the actual pairing symmetry is still under debate \cite{Wilson2024,Jiang2022}. 

First, we discuss the SC driven by fluctuations produced by V-V interactions, i.e. the $\gamma=0$ plane in Fig.~\ref{fig:Fig_3_main_text}(a). We see that fluctuations of V-V cBO or sLC lead to $s$-wave ($A_{1g}$) pairing with a large amplitude on the outer two FSs, but a small amplitude on the $\Gamma$-pocket, as shown in Fig.~\ref{fig:Fig_3_main_text} (d).  This is due to a small projection of the V-V channel interactions on the $\Gamma$-pocket. In Fig.~\ref{fig:Fig_3_main_text} (e) we see a qualitatively similar distribution of gaps across the FSs for the $d$-wave ($E_{2g}$) pairing driven by V-V cLC or sBO (see Figs.~\ref{fig:Fig_3_main_text} (a)). Upon including non-linear corrections to the gap equation, the two-fold degenerate $E_{2g}$ state can break time-reversal symmetry to become a chiral $d_{x^2-y^2}\pm i d_{xy}$ \cite{Sigrist1991}
to maximize condensation energy. 

On the other hand, as seen in Fig.~\ref{fig:Fig_3_main_text} (a), the V-Sb-ip channel fluctuations always mediate $s$-wave ($A_{1g}$) pairing. We find in Figs.~\ref{fig:Fig_3_main_text} (f,g) that fluctuating V-Sb-ip channels (close to $\gamma\approx 1$) induce a large gap across the $\Gamma$-pocket, in addition to comparably large gaps on the outer two FSs. This may be consistent with experiments that suggest sizable gaps on the $\Gamma$-pocket \cite{Deng2024}. We can further sub-classify the $A_{1g}$ region in Fig.~\ref{fig:Fig_3_main_text} (b), by the relative sign of the average pairing gap across the inner $\Gamma$-pocket (in) and the outer two FS (out). We quantify this by the ratio of angular summed gaps on the inner and outer FSs, i.e. $\Tilde{\Delta}_{\text{in}}/\Tilde{\Delta}_{\text{out}}=\oint _{\text{in}}\Delta_{\text{in}}(k) dk/\oint_{\text{out}} \Delta_{\text{out}}(k) dk$, and it is shown in Fig.~\ref{fig:Fig_3_main_text}(b).
Fluctuations of V-Sb-ip cLC or sBO channels create a gap with different signs across the outer two FS and inner $\Gamma$-pocket as shown in Fig.~\ref{fig:Fig_3_main_text} (g), resulting in $\Tilde{\Delta}_{\text{in}}/\Tilde{\Delta}_{\text{out}}<0$ in Fig.~\ref{fig:Fig_3_main_text} (b). This is due to a large repulsion in these channels between the outer V-dominated and inner-Sb-ip dominated FSs, leading to a $s_{+-}$  pairing. The opposite behavior is seen for the V-Sb-ip cBO or sLC channels which is always attractive, leading to an $s_{++}$ pairing as shown in Fig.~\ref{fig:Fig_3_main_text} (f), or $\Tilde{\Delta}_{\text{in}}/\Tilde{\Delta}_{\text{out}}>0$ in Fig.~\ref{fig:Fig_3_main_text} (b). 

\begin{figure}
    \centering
    \hspace{-0.5cm}\includegraphics[scale=0.8]{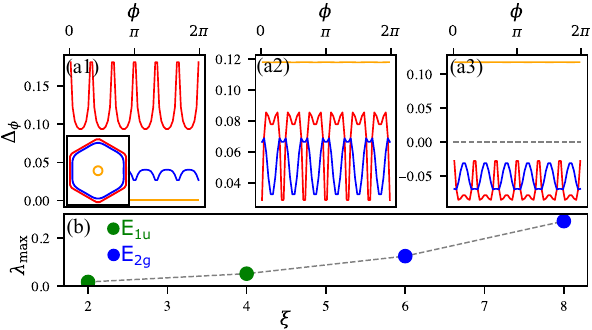}
    \caption{(a1-a3) The $s$-wave ($A_{1g}$) gaps $\Delta_{\phi}$ of Figs.~\ref{fig:Fig_3_main_text} (d, f, g), as a function of angle on the three Fermi sheets (see inset). (b) Triplet p-wave ($E_{1u}$) disappears on increasing $\xi$.}
    \label{fig:Fig_4_main_text}
\end{figure}

All the $s$-wave states in Fig.~\ref{fig:Fig_3_main_text} (a) have an anisotropic gap, whose angular distribution is shown in Fig.~\ref{fig:Fig_4_main_text} (a).
Further, in Figs.~\ref{fig:Fig_3_main_text} (a,c) we see a change in the pairing symmetry occurs on tuning $\alpha$ for a given mode (BOs or LCs), between $\alpha\sim 0.25-0.3$. This is because the spin vertex, with three components, contributes more than the charge vertex (details in Supplement \cite{supplement}). 
In Fig.~\ref{fig:Fig_3_main_text} (a), we also see a small region of a triplet $p$-wave ($E_{1u}$) for $\beta\approx 0.5$. However, this weak triplet phase, with small $\lambda_{\text{max}}$ values compared to the nearby singlet states, disappears on increasing the interaction scale $\xi$ as shown in Fig.~\ref{fig:Fig_4_main_text} (b).  Additionally, for each value of $\gamma$ we find that cLC and sBO fluctuations favor the same pairing symmetry (see Fig.~\ref{fig:Fig_3_main_text} (a)), as do cBO and sLC fluctuations.

\textit{Time-reversal parity of fluctuating mode and SC.-} 
The leading singlet pairing instability is closely tied to the time-reversal parity of the fluctuating modes \cite{Scheurer2016}. TRS even modes lead to conventional $s$-wave pairing, and TRS odd modes favor unconventional $s_{+-}$ or chiral $d+id$ pairing with sign-changing gaps. Since the BO \& LC channels each have definite TR parities, the pairing symmetry of the singlet states in Figs.~\ref{fig:Fig_3_main_text} (a,b) can be understood qualitatively based on  the TR parity of the strongest fluctuating mode. See End Matter for a more detailed discussion.

\textit{Discussions.--} In summary, by considering an electron model which incorporates nearest neighbor V-V and V-planar Sb interactions and the appropriate band structure effects, we have shown that $s_{++}$, $s_{+-}$, or $d+id$ superconductivity can arise from the competition between bond-order and loop-current fluctuations. 
All of these are consistent with experimental observations of a full gap in the SC state \cite{Deng_2024,Grant_2025,Roppongi2023,zhao2025,mine2024,Guguchia_2023} at low pressures.

Recent experiments probing the SC gap structure using electron-irradiation techniques \cite{Roppongi2023}, ARPES \cite{mine2024}, penetration depth measurements \cite{Grant_2025}, and point-contact Andreev-reflection spectroscopy (PCARS) \cite{zhao2025} have all reported the presence of highly anisotropic gaps. Further, the ARPES \cite{mine2024} observed the gap on the Sb-ip dominated $\Gamma$-pocket to be isotropic, with most of the anisotropy residing on the outer FSs. A similar coexistence of isotropic and anisotropic gaps was reported in the PCARS study \cite{zhao2025}. In the absence of TRS breaking, both the $s_{++}$ or $s_{+-}$ gaps in Figs.~\ref{fig:Fig_4_main_text} (a2,a3) can naturally explain this scenario. These states also have a large gap on the $\Gamma$-pocket (in the presence of a sufficiently large V-planar Sb interaction), consistent with quasi-particle interference spectroscopy measurements \cite{Deng2024}.
Distinguishing $s_{+-}$ and $s_{++}$ would require a carefully designed phase sensitive measurement. There have been various reports of time-reversal symmetry breaking \cite{Deng_2024,Gupta2022,Guguchia_2023}, which may favor $d+id$ state. However, this issue is still under debate as the impact of the extrinsic effects such as impurities and defects have yet to be clarified.

Recently, Schultz et.~al \cite{Schultz_2025} investigated the pairing mediated by charge-LC fluctuations between V-V and V-Sb-ip by incorporating a phenomenological LC propagator in the Cooper vertex, which introduced $M_{c}$-fluctuations without specifying its origin. Our results are consistent with these findings when we restrict our computations to the cLC sector.

\textit{Acknowledgements.--}
DJS and YBK thank G. Palle, R.M. Fernandes, J. Schmalian for earlier collaboration on a related subject and helpful discussions.
This work was supported by 
the Natural Science and Engineering Council of Canada (NSERC) and the Center for Quantum Materials at the University of Toronto. Computations were performed on the Cedar cluster hosted by WestGrid and SciNet in partnership with the Digital Research Alliance of Canada. This work was further supported by the German Research Foundation (DFG) through CRC TRR 288 ``Elasto-Q-Mat,'' project A07 (D.J.S.).
%\end{acknowledgements}

\twocolumngrid

\appendix

\section{END MATTER}
\subsection{Superconducting Instabilities}
To include the effect of the different fluctuating modes (present across all $\mathbf{q}$) in the interaction vertex, we compute the RPA-corrected vertex: $V=V_0+V_{\text{RPA}}$, by summing over all bubble-diagrams (more details in \cite{supplement}).
\begin{gather}
\left[V_{\text{RPA}}(\mathbf{k},\mathbf{k'})\right]^{\Tilde{\mu}_1\Tilde{\mu}_2}_{\Tilde{\mu}_3\Tilde{\mu}_4}=-\left[ V_0 (1+\chi_0 V_0)^{-1} \chi_0 V_0 (\mathbf{k},\mathbf{k'})\right]^{\Tilde{\mu}_1\Tilde{\mu}_2}_{\Tilde{\mu}_3\Tilde{\mu}_4},\label{eq:RPA_correction_to_V}
\end{gather}
where $\Tilde{\mu}=(\mu,s)$ is a combined orbital-spin index. Eq.~\eqref{eq:RPA_correction_to_V} includes a static orbital-resolved susceptibility $\chi_0(\mathbf{q},\omega=0)$ (defined in \cite{supplement}) and no dynamical effects to the pairing are considered. Eq.~\eqref{eq:RPA_correction_to_V} is then projected onto the Cooper channel and anti-symmetrized. This results in an effective pairing interaction of the form,
\begin{gather}
\hspace{-0.17cm}H_{\text{eff}}=\frac{1}{2N}\sum_{\mathbf{k},\mathbf{k'},\{\Tilde{\mu}\}}V(\mathbf{k},\mathbf{k'})^{\Tilde{\mu}_1\Tilde{\mu}_2}_{\Tilde{\mu}_3\Tilde{\mu}_4}c_{\mathbf{k}\Tilde{\mu}_1}^{\dagger}c_{\mathbf{-k}\Tilde{\mu}_3}^{\dagger}
    c_{-\mathbf{k'}\Tilde{\mu}_2}c_{\mathbf{k'}\Tilde{\mu}_4}.\label{eq:Cooper_1_mt}
\end{gather}
We find the leading and sub-leading SC instabilities, with intra-band pairing, by solving a linearized gap equation for the Cooper vertex in the singlet $(\eta=0)$ and triplet ($\eta=\{x,y,z\}$) 
channels,
\begin{gather}
\hspace{-0.05cm}\lambda_{\eta}\Delta_{\eta}(n_1\mathbf{k})= - \hspace{-0.1cm}\sum_{n_2}\oint_{\mathbf{k'}\in \text{FS}_{n_2}}\hspace{-0.7cm} \frac{\Tilde{V}_{\eta}(n_1\mathbf{k},n_2\mathbf{k'}) \Delta_{\eta} (n_2\mathbf{k'}) d\mathbf{k'}} {\left|\nabla_{\mathbf{k'}}\xi_{\mathbf{k'}n_2}\right| \left(2\pi \right)^2}. \label{eq:LGE}
\end{gather}
Here, $n\mathbf{k}$ denotes a FS momenta $\mathbf{k}$ of band $n$. The triplet channels are degenerate because of unbroken spin-rotation symmetry in the disordered state and the absence of spin-orbit coupling. The eigenvector $\Delta_{\text{max}}(n\mathbf{k})$ of the leading SC instability with the largest eigenvalue $\lambda_{\text{max}}$, encodes its gap structure across the FS. We classify the gap functions based on lattice harmonics and irreps of the $D_{6h}$ point group. Since in this study we primarily focus on the competition between different fluctuating channels, we set the overall interaction scale $\xi=4$. For reference, the average band-width of the V-$d$ bands near the VHS is $W_{\text{band}}\sim 12$ (here, we set $t_{\text{V-V}}/2=1$, see \cite{supplement}). Small changes to $\xi$ do not qualitatively change the leading SC instability (see Fig.~\ref{fig:Fig_3_main_text} (c)). All our results are obtained at $T/t_{\text{V-V}}\sim 0.005$, representative of the low-temperature limit.
\\
\subsection{Time-reversal parity of fluctuating mode and SC}
The singlet-channel Cooper vertex for a fluctuating mode of TR parity $P_T$ can be expressed as
\begin{gather}
\hspace{-0.1cm}\Tilde{V}_{\eta=0}(n_1\mathbf{k},n_2\mathbf{k'})=\hspace{-0.1cm}\frac{-g^2P_T}{2}\left( \left|\mathcal{O}^{n_1,n_2}_{\mathbf{k},\mathbf{k'}}\right|^2 +\left|\mathcal{O}^{n_1,n_2}_{\mathbf{k},\mathbf{-k'}}\right|^2 \right)\label{eq:main_singlet_parity},
\end{gather}
where $\mathcal{O}^{n_1,n_2}_{\mathbf{k},\mathbf{k'}}$ is the fluctuating mode in the band-basis (a simple proof presented in \cite{supplement}). Eq.~\eqref{eq:main_singlet_parity} shows that the $P_T = 1$ singlet channels are attractive ($\tilde{V}_{\eta=0}(\mathbf{k},\mathbf{k}') < 0$) across the entire FS, leading to conventional $s$-wave pairing. $P_T = -1$ modes are repulsive ($\tilde{V}_{\eta=0}(\mathbf{k},\mathbf{k}') > 0$), therefore favoring unconventional sign-changing gaps. The BO \& LC channels each have definite TR parities and therefore the pairing symmetry of the singlet states in Figs.~\ref{fig:Fig_3_main_text} (a,b) can be understood qualitatively based on  the TR parity of the strongest fluctuating mode. 
For example, focusing on the charge sector ($\alpha=0$) in Fig.~\ref{fig:Fig_3_main_text} (a), dominant TRS even cBO fluctuations (at large $\beta$)
drive $s_{++}$-wave pairing. On the other hand, dominant cLC fluctuations (at small $\beta$) drive unconventional chiral $d+id$ or $s_{+-}$ pairing. These conclusions remain valid after incorporating RPA corrections to Eq.~\eqref{eq:main_singlet_parity}. However, this argument alone does not specify whether the leading pairing state is singlet or triplet.

\end{document}

\title{\textit{Supplement to}: Interplay of competing bond-order and loop-current fluctuations as a possible
mechanism for superconductivity in kagome metals}

\author{Asimpunya Mitra$^{1}$}
\author{Daniel J. Schultz$^{2}$}
\author{Yong Baek Kim$^{1}$}

\affiliation{$^1$Department of Physics, University of Toronto, Toronto, Ontario M5S 1A7, Canada}
\affiliation{$^2$Institute for Theoretical Condensed Matter Physics,
Karlsruhe Institute of Technology, 76131 Karlsruhe, Germany}
\date{\today}   

\maketitle

In this supplement to the main text, we have provided more details about the calculations and results. In Sec. \ref{sec:lattice_conventions_supp}, we introduce the 8-band tight-binding model, and briefly discuss the consequences of mirror symmetries of the VHS. In Sec. \ref{sec:vertex_LC_BO_supp}, starting from the nearest-neighbor Coulomb interaction, we discuss its splitting into the different bond-order and loop-current channels. In Sec. \ref{sec:vertex_Cooper_supp}, we discuss the RPA-corrected susceptibilities and RPA-corrected Cooper vertex, and its transformation into band-space and singlet/triplet sectors. In Sec. \ref{sec:LGE_supp}, we mention the linearized gap equation, and the classification of the gap functions. In Sec. \ref{sec:T_parity_supp}, we give a simple proof of the fact that fluctuations of a time-reversal even/odd mode give rise to a conventional/unconventional gap function in the singlet channel, and we discuss the consequences of adding RPA-corrections on this result. In Sec. \ref{sec:visualizing_vertices}, we plot the intensities of the singlet and triplet Cooper vertices for different fluctuating channels. In Sec. \ref{sec:5.02}, we present additional results for $\mu=5.02$, i.e. when the chemical potential is near the M-type VHS.

\section{Electronic Structure}\label{sec:lattice_conventions_supp}
\subsection{The 8-band tight binding model}\label{sec:1a}
\begin{figure}
    \centering
    \includegraphics[scale=0.75]{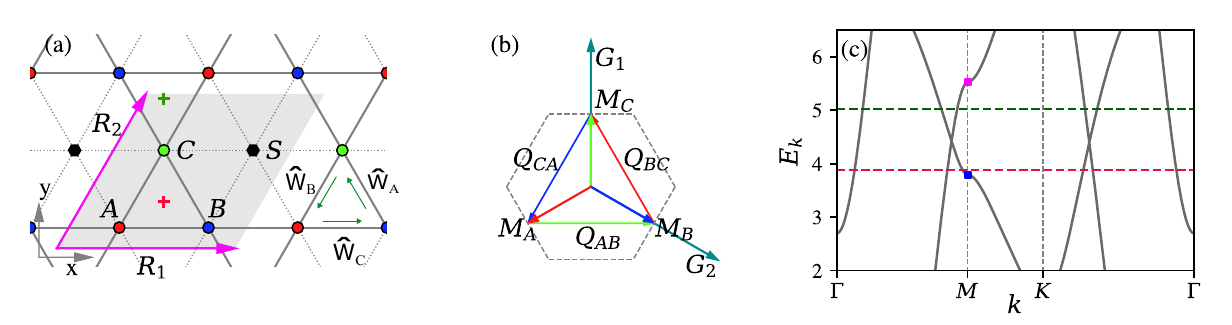}
    \caption{(a) The 2D structure of \ce{AV3Sb5}. The unit cell is shaded grey. The green arrows $\hat{\text{w}}_{A/B/C}$ show the different bond directions. (b) The corresponding Brillouin zone. (c) The electronic band structure of the 8-band model. The band structure features P-type VHS (red) and M-type VHS (blue), and an Fermi-surface pocket near $\Gamma$. The chemical potentials $\mu=3.88$ ($=5.02$) are marked with the red (green) dashed lines. We define, $\mathbf{x}_C=\mathbf{R}_1$, $\mathbf{x}_A=\mathbf{R}_2-\mathbf{R}_1$, $\mathbf{x}_B=-\mathbf{R}_2$.}
    \label{fig:1_supp}
\end{figure}

The kagome metals AV${}_3$Sb${}_5$ (A=K, Rb, Cs) have a  quasi-two dimensional layered structure. The vanadium (V) ions form the kagome net (see Fig.~\ref{fig:1_supp} (a)). There are two kinds of antimony (Sb) positions. The in-plane-Sb (Sb-ip) are present at the center of the hexagons of the kagome lattice ($S$ positions in Fig.~\ref{fig:1_supp} (a)), the out-of-plane-Sb (Sb-op) are present above and below the kagome plane (at the two $+$ positions in the 2D projection in Fig.~\ref{fig:1_supp} (a)). Due to the layered structure of these \ce{AV3Sb5} kagome metals, we focused on a single kagome plane and analyzed only the $k_z=0$ states, and constructed effective 2D tight-binding models to study these systems. 

The kagome lattice has the Bravais lattice vectors: $\boldsymbol{R_1}=(1,0)$, $\boldsymbol{R_2}=(\frac{1}{2},\frac{\sqrt{3}}{2})$. The three sub-lattice positions called $A,B,C$ (Fig.~\ref{fig:1_supp} (a)) have positions $\boldsymbol{\delta_A}= (0,0)$, $\boldsymbol{\delta_B}=(\frac{1}{2},0)$, $\boldsymbol{\delta_C}=(\frac{1}{4},\frac{\sqrt{3}}{4})$ respectively. The reciprocal lattice vectors are $\boldsymbol{G_1}=2\pi(0,\frac{2}{\sqrt{3}})$, $\boldsymbol{G_2}=2\pi(1,-\frac{1}{\sqrt{3}})$. Additionally, from Fig.~\ref{fig:1_supp} (a), we see that $Q_{AB}\equiv M_C$, $Q_{BC}\equiv M_A$, $Q_{CA}\equiv M_B$.

We constructed an 8-band tight-binding model to describe the electronic structure of \ce{AV3Sb5} close to the Fermi energy, which in actual materials resides between the P-type and M-type VHS as shown in Fig.~\ref{fig:1_supp} (c). The goal of the model was to: (i) capture the correct nature of the P- and M-type van-Hove singularities (VHS) near the Fermi surface, with the P-type mostly formed from the V-$3d$ orbitals, and M-type being formed as a result of strong hybridization between V and Sb-op orbitals \cite{Jeong_2022,Li_2023}. (ii) Ensure that the orbitals considered are mirror-odd ($\sigma_h=-1$), and the wavefunctions at the VHS have the mirror symmetries described in \cite{Li_2023,Li_2024} (brief summary in Sec. \ref{mirror_symm}). (iii) Include the Fermi surface pocket around $\Gamma$ which is mostly formed from the Sb-ip orbitals. (iv)  Only include nearest-neighbor hopping for simplicity. The 8-band model is constructed based on the DFT band structure in \cite{Li_2023}. However, we had to significantly deviate from some of the DFT parameters to ensure the above requirements are satisfied.

In the 8-band model, the vanadium in sub-lattices $A,B,C$ contributes one effective V-$3d$ orbital per V site to the 8-band model. These effective V-$d$ orbitals are called the $\Tilde{d}_{A},\Tilde{d}_{B},\Tilde{d}_{C}$ orbitals (one $\Tilde{d}$ orbital V per-site). These $\Tilde{d}$ orbitals are a linear combination of the $d_{xz}$ and $d_{yz}$ orbitals present at each V site,  and are defined below. The Sb-ip is accounted for by the Sb-$5p_{z}$ orbital. Additionally, we consider contributions from four Sb-op orbitals in the unit cell, which are anti-symmetric combinations of the Sb-$5p_{x}$ and Sb-$5p_{y}$ above and below the kagome plane (belonging to the $\sigma_h=-1$ sub-sector). Therefore, the orbitals that are considered in the 8-band model are:
\begin{gather}
    \Tilde{d}_{A}=-\frac{1}{2}d_{A,xz}+\frac{\sqrt{3}}{2}d_{A,yz}, \quad \Tilde{d}_{B}=-\frac{1}{2}d_{B,xz}-\frac{\sqrt{3}}{2}d_{B,yz} ,\quad \Tilde{d}_{C}=d_{C,xz},\quad p_{z}=p_{\text{Sb}_{\text{ip}},\text{z}}\nonumber\\
    p_{1x}=\frac{1}{\sqrt{2}}\left(p_{1x,\text{Sb}_{\text{op}}}^{\text{up}}-p_{1x,\text{Sb}_{\text{op}}}^{\text{down}} \right),\quad
    p_{1y}=\frac{1}{\sqrt{2}}\left(p_{1y,\text{Sb}_{\text{op}}}^{\text{up}}-p_{1y,\text{Sb}_{\text{op}}}^{\text{down}} \right),\nonumber\\
    p_{2x}=\frac{1}{\sqrt{2}}\left(p_{2x,\text{Sb}_{\text{op}}}^{\text{up}}-p_{2x,\text{Sb}_{\text{op}}}^{\text{down}} \right),\quad 
    p_{2y}=\frac{1}{\sqrt{2}}\left(p_{2y,\text{Sb}_{\text{op}}}^{\text{up}}-p_{2y,\text{Sb}_{\text{op}}}^{\text{down}} \right),
\end{gather}
\begin{figure}
    \centering
    \includegraphics[scale=0.9]{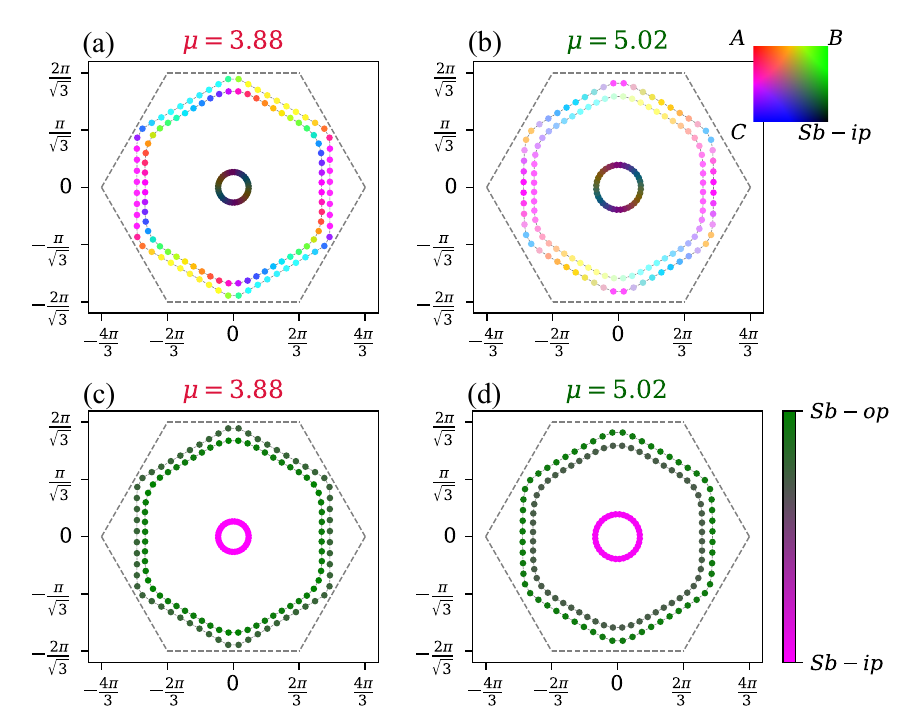}
    \caption{(a) and (b) show the 3-sheet Fermi-surface at chemical potentials $\mu=3.88$ (near P-type VHS), $\mu=5.02$ (near M-type VHS) respectively. The four-point color-scale shows the weight of the V-$\Tilde{d}$ orbitals at at sub-lattices $A$, $B$, $C$, and the Sb-$p_z$ orbital from the in-plane-Sb (Sb-ip). The effect of the sub-lattice interference near the P-and M- type VHS can be traced from the color scale in (a) and (b). (c) and (d) show the weight of the Sb-ip and out-of-plane Sb (Sb-op) orbitals across the Fermi surface. As shown in (d), for $\mu=5.02$ (close to M-type VHS) the outer most Fermi surface has a comparatively large weight from the Sb-op orbitals which is essential for forming the M-type VHS.}
    \label{fig:2_supp}
\end{figure}
where $p_{1}$ and $p_{2}$ denote the positions of the Sb-op present in the unit cell on the lower $+$ (in red in Fig.~\ref{fig:1_supp} (a)) and upper $+$ (in green in Fig.~\ref{fig:1_supp} (a)) positions respectively in the unit cell as shown in Fig.~\ref{fig:1_supp} (a). Further up, down denote the Sb-op positions above and below the kagome plane. The 8-band tight-binding Hamiltonian is
\begin{gather}
H_0=\sum_{\mathbf{k}}\Psi_{\mathbf{k}\mu_1\sigma_1}^{\dagger}\left(\mathds{1}_{\sigma_1\sigma_2}\otimes \left[H\left(\mathbf{k}\right)\right]_{\mu_1\mu_2}\right)\Psi_{\mathbf{k}\mu_2\sigma_2}\label{eq:ham_def_supp}
\end{gather}
where we define the vector $\Psi_{\mathbf{k}\mu\sigma}^{\dagger}=\left(\Tilde{d}_{\mathbf{k}A\sigma}^{\dagger},\, \Tilde{d}_{\mathbf{k}B\sigma}^{\dagger},\,\Tilde{d}_{\mathbf{k}C\sigma}^{\dagger},\,
p_{\mathbf{k}1x\sigma}^{\dagger},\,
p_{\mathbf{k}1y\sigma}^{\dagger},\,
p_{\mathbf{k}2x\sigma}^{\dagger},\,
p_{\mathbf{k}2y\sigma}^{\dagger},\,
p_{\mathbf{k}z\sigma}^{\dagger}
\right)^T$. The orbital portion of Eq.~\eqref{eq:ham_def_supp} is composed of different blocks
\begin{gather}
    H\left(\mathbf{k}\right)=
    \begin{pmatrix}
        \left[H_{\text{V-V}}(\mathbf{k}) \right]_{3\times 3}   & \left[H_{\text{V-Sb}_{\text{op}}}(\mathbf{k}) \right]_{3\times 4} & \left[H_{\text{V-Sb}_{\text{ip}}}(\mathbf{k}) \right]_{3\times 1}\\
        \left[H_{\text{V-Sb}_{\text{op}}}(\mathbf{k}) \right]_{4\times 3}^{\dagger} & \left[H_{Sb_{op}-Sb_{op}}(\mathbf{k}) \right]_{4\times 4} & \left[H_{\text{Sb}_{\text{op}}-\text{Sb}_{\text{ip}}}(\mathbf{k}) \right]_{4\times 1}\\
        \left[H_{\text{V-Sb}_{\text{ip}}}(\mathbf{k}) \right]_{1\times 3}^{\dagger} & \left[H_{\text{Sb}_{\text{op}}-\text{Sb}_{\text{ip}}}(\mathbf{k}) \right]_{1\times 4}^{\dagger} & \left[H_{\text{Sb}_{\text{ip}}-\text{Sb}_{\text{ip}}}(\mathbf{k}) \right]_{1\times 1}
    \end{pmatrix}
\end{gather}
The V-V block is the usual tight-binding for the kagome lattice with $t_{\text{V-V}}=2.0$, $\epsilon_{\text{V}}=0$,
\begin{gather}
    H_{\text{V-V}}(\mathbf{k})=
    \begin{pmatrix}
    \epsilon_{\text{V}} & t_{\text{V-V}}\left(1+e^{-i \mathbf{k}\cdot \mathbf{R_1}}\right) & t_{\text{V-V}}\left(1+e^{-i \mathbf{k}\cdot \mathbf{R_2}}\right) \\
    t_{\text{V-V}}\left(1+e^{i \mathbf{k}\cdot \mathbf{R_1}}\right) & \epsilon_{\text{V}} & t_{\text{V-V}}\left(1+e^{-i \mathbf{k}\cdot (\mathbf{R_2}-\mathbf{R_1})}\right) \\
    t_{\text{V-V}}\left(1+e^{i \mathbf{k}\cdot \mathbf{R_2}}\right) & t_{\text{V-V}}\left(1+e^{i \mathbf{k}\cdot (\mathbf{R_2}-\mathbf{R_1})}\right) & \epsilon_{\text{V}}
    \end{pmatrix}.
    \label{eq:H_(V,V)}
\end{gather}
The Sb-op hoppings are 
$t_{\text{Sb}_{\text{op}}-\text{Sb}_{\text{op}}}^{(\sigma)}=13.0$,
$t_{\text{Sb}_{\text{op}}-\text{Sb}_{\text{op}}}^{(\pi)}=-2.0$,  $\epsilon_{\text{Sb}_{\text{op}}}=2.0$
\begin{gather}
H_{Sb_{op}-Sb_{op}}(\mathbf{k})=\begin{pmatrix}
    \epsilon_{\text{Sb}_{\text{op}}} & 0 & H_{0,2}(\mathbf{k}) & H_{0,3}(\mathbf{k})\\
    0 & \epsilon_{\text{Sb}_{\text{op}}} &  H_{1,2}(\mathbf{k}) & H_{1,3}(\mathbf{k}) \\
    H_{0,2}^{*}(\mathbf{k})& H_{1,2}^{*}(\mathbf{k})& \epsilon_{\text{Sb}_{\text{op}}} & 0\\
    H_{0,3}^{*}(\mathbf{k})& H_{1,3}^{*}(\mathbf{k})& 0 & \epsilon_{\text{Sb}_{\text{op}}}
    \end{pmatrix},
    \label{eq:H_(Sbop,Sbop)}
\end{gather}
where
\begin{align}
    H_{0,2}(\mathbf{k})=& t^{(\pi)}_{\text{Sb}_{\text{op}}-\text{Sb}_{\text{op}}}+\left(\frac{3}{4}t^{(\sigma)}_{\text{Sb}_{\text{op}}-\text{Sb}_{\text{op}}}+\frac{1}{4}t^{(\pi)}_{\text{Sb}_{\text{op}}-\text{Sb}_{\text{op}}}\right)\left(e^{-i\mathbf{k}\cdot \mathbf{R_2}}+e^{-i\mathbf{k}\cdot\left( \mathbf{R_2}-\mathbf{R_1}\right)}\right),\\
    H_{0,3}(\mathbf{k})=&\frac{\sqrt{3}}{4}\left(t^{(\sigma)}_{\text{Sb}_{\text{op}}-\text{Sb}_{\text{op}}}-t^{(\pi)}_{\text{Sb}_{\text{op}}-\text{Sb}_{\text{op}}}\right)\left( e^{-i\mathbf{k}\cdot \mathbf{R_2}}-e^{-i\mathbf{k}\cdot\left( \mathbf{R_2}-\mathbf{R_1}\right)}\right),\\
     H_{1,2}(\mathbf{k})=&\frac{\sqrt{3}}{4}\left(t^{(\sigma)}_{\text{Sb}_{\text{op}}-\text{Sb}_{\text{op}}}-t^{(\pi)}_{\text{Sb}_{\text{op}}-\text{Sb}_{\text{op}}}\right)\left( e^{-i\mathbf{k}\cdot \mathbf{R_2}}-e^{-i\mathbf{k}\cdot\left( \mathbf{R_2}-\mathbf{R_1}\right)}\right),\\
     H_{1,3}(\mathbf{k})=& t^{(\sigma)}_{\text{Sb}_{\text{op}}-\text{Sb}_{\text{op}}}+\left(\frac{3}{4}t^{(\pi)}_{\text{Sb}_{\text{op}}-\text{Sb}_{\text{op}}}+\frac{1}{4}t^{(\sigma)}_{\text{Sb}_{\text{op}}-\text{Sb}_{\text{op}}}\right)\left(e^{-i\mathbf{k}\cdot \mathbf{R_2}}+e^{-i\mathbf{k}\cdot\left( \mathbf{R_2}-\mathbf{R_1}\right)}\right).
\end{align}
To stabilize the Sb-ip dominated $\Gamma$-pocket in the Fermi surface we introduced a hopping between the Sb-ip sites, where $t_{\text{Sb}_{\text{ip}}-\text{Sb}_{\text{ip}}}=1.8$, $\epsilon_{\text{Sb}_{\text{ip}}}=13.5$
\begin{gather}
    H_{\text{Sb}_{\text{ip}}-\text{Sb}_{\text{ip}}}(\mathbf{k})=\epsilon_{\text{Sb}_{\text{ip}}}+2t_{\text{Sb}_{\text{ip}}-\text{Sb}_{\text{ip}}}\left(
    \cos \left(\mathbf{k}\cdot \mathbf{R_1}\right)
    +\cos \left(\mathbf{k}\cdot \mathbf{R_2}\right) 
    +\cos \left(\mathbf{k}\cdot \left(\mathbf{R_2}-\mathbf{R_1}\right) \right)
    \right).
    \label{eq:H_(Sbip,Sbip)}
\end{gather}
The hopping between the V and Sbop is given by $t_{\text{V-Sb}_{\text{op}}}=6.5$,
\begin{gather}
    H_{\text{V-Sb}_{\text{op}}}(\mathbf{k})=
    t_{\text{V-Sb}_{\text{op}}}\begin{pmatrix}
       \frac{1}{2} & -\frac{\sqrt{3}}{2}& \frac{1}{2} e^{-i \mathbf{k}\cdot \mathbf{R_2}} & -\frac{\sqrt{3}}{2} e^{-i \mathbf{k}\cdot \mathbf{R_2}}\\
        \frac{1}{2} & \frac{\sqrt{3}}{2} & \frac{1}{2}e^{-i \mathbf{k}\cdot \left(\mathbf{R_2}-\mathbf{R_1}\right)}& \frac{\sqrt{3}}{2}e^{-i \mathbf{k}\cdot \left(\mathbf{R_2}-\mathbf{R_1}\right)}\\
        -1 & 0 & -1 & 0
    \end{pmatrix}.
    \label{eq:H_(V,Sbop)}
\end{gather}
The hopping between the V and Sbip is given by $t_{\text{V-Sb}_{\text{ip}}}=3.5$,
\begin{gather}
    H_{\text{V-Sb}_{\text{ip}}}(\mathbf{k})=t_{\text{V-Sb}_{\text{ip}}}\begin{pmatrix}
       1-e^{i\mathbf{k}\cdot \left(\mathbf{R_2}-\mathbf{R_1}\right)}\\
       -1+e^{i\mathbf{k}\cdot \mathbf{R_2}}\\
       -e^{i\mathbf{k}\cdot \mathbf{R_2}}+e^{i\mathbf{k}\cdot \left(\mathbf{R_2}-\mathbf{R_1}\right)}
    \end{pmatrix}.
    \label{eq:H_(V,Sbip)}
\end{gather}
The hopping between the Sb-op and Sbip is $t_{\text{Sb}_{\text{op}}-\text{Sb}_{\text{ip}}}=5.0$,
\begin{gather}
    H_{\text{Sb}_{\text{op}}-\text{Sb}_{\text{ip}}}(\mathbf{k})=t_{\text{Sb}_{\text{op}}-\text{Sb}_{\text{ip}}}\begin{pmatrix}
        \frac{\sqrt{3}}{2}\left(e^{i\mathbf{k}\cdot \left(\mathbf{R_2}-\mathbf{R_1}\right)}- e^{i\mathbf{k}\cdot\mathbf{R_2}}\right)\\
        1-\frac{1}{2}\left(e^{i\mathbf{k}\cdot \left(\mathbf{R_2}-\mathbf{R_1}\right)}+ e^{i\mathbf{k}\cdot\mathbf{R_2}}\right)\\
        \frac{\sqrt{3}}{2}\left(e^{i\mathbf{k}\cdot \left(\mathbf{R_2}-\mathbf{R_1}\right)}- e^{i\mathbf{k}\cdot\mathbf{R_2}}\right)\\
        \frac{1}{2}\left(e^{i\mathbf{k}\cdot \left(\mathbf{R_2}-\mathbf{R_1}\right)}+ e^{i\mathbf{k}\cdot\mathbf{R_2}}\right)- e^{i\mathbf{k}\cdot \left(\mathbf{2R_2}-\mathbf{R_1}\right)}
    \end{pmatrix}.
    \label{eq:H_(V,Sbip)}
\end{gather}

Near the P- and M-type VHS, the 8-band tight model has a dispersion as shown in Fig.~\ref{fig:1_supp} (c). The P-type VHS is mostly composed of the V-$\Tilde{d}$ orbitals, and the M-type VHS is made from the V-$\Tilde{d}$ and the Sb-op $p$ orbitals. The $\Gamma$-pocket is mostly formed from the Sb-ip $p_{\text{z}}$ orbital. These features are shown in Fig.~\ref{fig:2_supp}. Further, the wavefunctions at the VHSs have the mirror symmetries described in \cite{Li_2023} (also discussed in Sec. \ref{mirror_symm}). Mean-field calculations show that this model hosts both bond-orders and loop-currents (this will feature in a future study).

\subsection{P- and M-type Van-Hove singularities}\label{mirror_symm}
The VHS near the Fermi level coming from the V-$\Tilde{d}$ orbitals are classified based on the contribution of (the orbitals on) the different V sub-lattices to the wave function. The P-type (sub-lattice pure) VHS has only one V sub-lattice contributing in its wave-function \cite{Kiesel_2012}. For example, the P-type VHS at $\mathbf{q}=M_{C}$ has contribution from only the orbitals on the $C$ sub-lattice (or $\Tilde{d}_{C}$ orbital), and its wavefunction has the form $\Psi_{P,C}=(0,0,a)$ where the three components specify the contribution from the orbitals on the three sub-lattices-$(A,B,C)$. In contrast, the M-type (mixed type) VHS, at the same momentum has contributions from (the orbitals on) sub-lattices $A$ \& $B$, and the wave-function can have the form $\Psi_{M+,C}=(b,b,0)$ or $\Psi_{M-,C}=(b,-b,0)$. In our 8-band model we have the M-type VHS of the form $\Psi_{M-,C}=(b,-b,0)$, which we say has the correct \textit{mirror} symmetry based on the terminology from Ref. \cite{Li_2024}. It has been shown in Ref. \cite{Li_2024} that, the presence of the P-type and the M-type VHS with this correct \textit{mirror} symmetry, with the Fermi energy in between them, enables a symmetry-allowed interaction between the VHSs, which can lower the free energy of the loop-current order. Since we study the superconductivity (SC) in the disordered state, we expect the tendency of the \textit{mirror} symmetries to stabilize the LC order to cause enhancement of the LC fluctuations in the disordered state.

\section{Constructing the Interaction Vertex}\label{sec:vertex_LC_BO_supp}
\subsection{Microscopic interaction}

\begin{figure}
    \centering
    \includegraphics[scale=0.75]{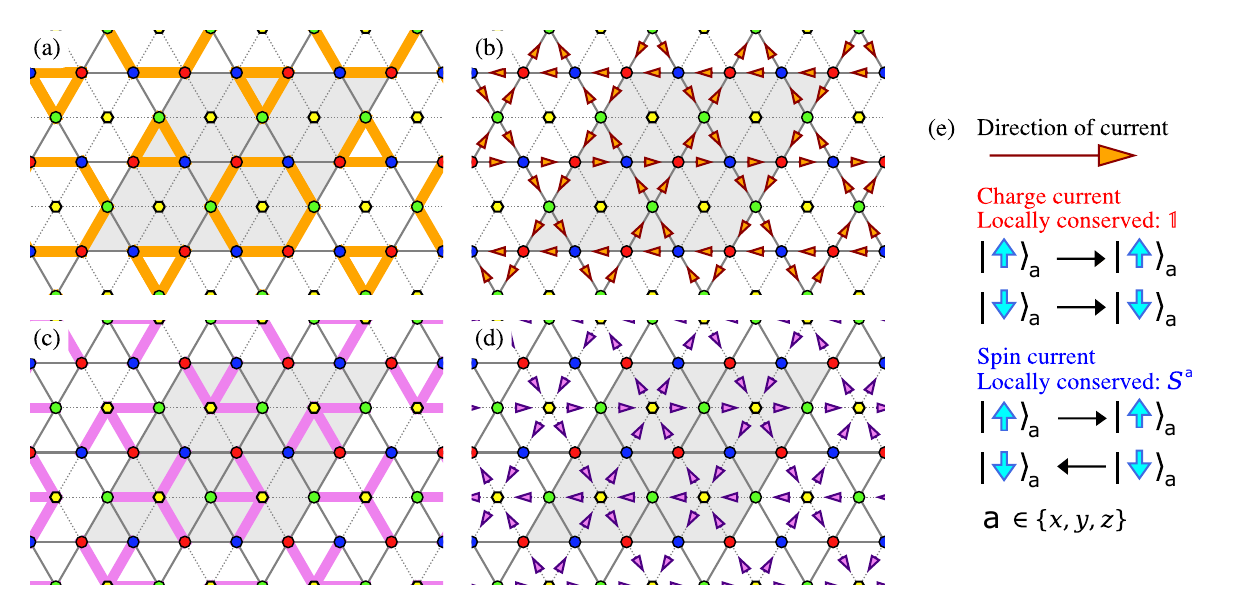}
    \caption{The ordered states resulting from condensation of the three $\mathbf{q}=M_{c}$ (or $3\mathbf{Q}$ states)  for each BO \& LC channel is shown in (a-d). The pattern of condensed (a) V-V bond-order, (b) V-V loop-current, (c) V-Sbip bond-order, (d) V-Sb-ip loop-current. The patterns shown here have three ordering wave-vectors $M_A$, $M_B$, $M_C$, and the order parameter has the form, \textit{e.g.} $\left(\mathcal{B}^{\textit{i},\text{vv}}_{M_A,A},\mathcal{B}^{\textit{i},\text{vv}}_{M_B,B},\mathcal{B}^{\textit{i},\text{vv}}_{M_C,C} \right)$ in (a). (e) Schematic showing the difference between charge and spin currents.}
    \label{fig:3_supp}
\end{figure}
In this study, we only consider the effect of the nearest-neighbor (NN) Coulomb repulsions $\mathcal{V}$. We do not consider the on-site Hubbard interaction, as it is suppressed close to the P-type VHS, due to sub-lattice interference \cite{Kiesel_2012,Kiesel_2013}. In addition to the usual NN Coulomb repulsion between the V sites $\mathcal{V}_{\text{vv}}$, we also consider the effect of the NN Coulomb between the V and Sb-ip sites $\mathcal{V}_{\text{sv}}$. This choice is motivated by a recent study \cite{Schultz_2025}, which phenomenologically analyzed the effect of fluctuating V-Sb-ip loop-currents in mediating superconductivity. The interaction Hamiltonian $H_{\text{int}}$ is 
\begin{gather}
H_{\text{int}}
=\mathcal{V}_{\text{vv}} \sum_{\mathcal{\mathbf{r}}_j,(a,b,c)} \left(n^{\text{v}}_{\mathcal{\mathbf{r}}_j,a} n^{\text{v}}_{\mathcal{\mathbf{r}}_j,b}+n^{\text{v}}_{\mathcal{\mathbf{r}}_j,a} n^{\text{v}}_{\mathcal{\mathbf{r}}_j-\mathbf{x}_{c},b}\right)
+\mathcal{V}_{\text{sv}} \sum_{\mathcal{\mathbf{r}}_j,(a,b,c)} n^{\text{s}}_{\mathcal{\mathbf{r}}_j+\mathbf{\Tilde{x}}_{a}} \left(n^{\text{v}}_{\mathcal{\mathbf{r}}_j,b}+ n^{\text{v}}_{\mathcal{\mathbf{r}}_j,c} \right) \label{eq:NN_Coulomb_interacting_ham}
\end{gather}
where $(a,b,c)$ denotes cyclic permutations of the $(A,B,C)$ and 
$(\mathbf{{\Tilde{x}}}_A, \mathbf{{\Tilde{x}}}_B, \mathbf{{\Tilde{x}}}_C) = 
(0, -\mathbf{x}_C, \mathbf{x}_B)$. $\mathbf{x}_A, \mathbf{x}_B, \mathbf{x}_C$ are defined in Fig.\ref{fig:1_supp} caption. 
In Eq.~\eqref{eq:NN_Coulomb_interacting_ham},  $\mathbf{r}_j$ labels the unit cell, and $n^{\text{v}}_{\mathbf{r}_j,a}$ denotes the density of the $V$ at sub-lattice $a$ in the unit cell, and $n^{\text{s}}_{\mathbf{r}_j}$ denotes the density at the Sb-ip site. The local density of the V at sub-lattice $a$, or the Sb-ip (s), can be expressed as
\begin{gather}
n^{\text{v}}_{\mathbf{r}_j,a}=\sum_{\mathcal{O}(a)}c^{\dagger}_{\mathbf{r}_j,a,\mathcal{O}(a)}c_{\mathbf{r}_j,a,\mathcal{O}(a)}=c^{\dagger}_{\mathbf{r}_j,a}c_{\mathbf{r}_j,a}\label{eq:d1}\\
n^{\text{s}}_{\mathbf{r}_j}=\sum_{\mathcal{O}(s)}c^{\dagger}_{\mathbf{r}_j,s,\mathcal{O}(s)}c_{\mathbf{r}_j,s,\mathcal{O}(s)}=c^{\dagger}_{\mathbf{r}_j,s}c_{\mathbf{r}_j,s}\label{eq:d2}
\end{gather}
In the above Eqs.~\eqref{eq:d1},~\eqref{eq:d2}, $\mathcal{O}(a)/\mathcal{O}(s)$ denotes the orbitals at the local sub-lattice positions. However, in our 8-band model we have one V-$\Tilde{d}$ orbital per V site (at V sub-lattice $a$, we have the $\Tilde{d}_a$ orbital), and also one Sb-$p_z$ orbital per Sb-ip site. Therefore, using this fact we write down the last equality of Eqs.~\eqref{eq:d1},~\eqref{eq:d2}, and do not keep the redundant $\mathcal{O}(a)/\mathcal{O}(s)$ index. 
Henceforth, we use a compact notation to label the fermion operators, and make the following identifications compared to Sec. \ref{sec:1a},
\begin{gather}
\Tilde{d}_{\mathbf{k}A\sigma}\rightarrow c_{\mathbf{k}A\sigma},\, \Tilde{d}_{\mathbf{k}B\sigma}\rightarrow c_{\mathbf{k}B\sigma},\,\Tilde{d}_{\mathbf{k}C\sigma}\rightarrow c_{\mathbf{k}C\sigma},\,
p_{\mathbf{k}1x\sigma}\rightarrow c_{\mathbf{k}1x\sigma},\\
p_{\mathbf{k}1y\sigma}\rightarrow c_{\mathbf{k}1y\sigma},\,
p_{\mathbf{k}2x\sigma}\rightarrow c_{\mathbf{k}2x\sigma},\,
p_{\mathbf{k}2y\sigma}\rightarrow c_{\mathbf{k}2y\sigma},\,
p_{\mathbf{k}z\sigma}\rightarrow c_{\mathbf{k}s\sigma}.
\end{gather}
Using this notation, we can express $H_{\text{int}}$ in momentum space as
\begin{align}
H_{\text{int}}=-&\frac{\mathcal{V}_{\text{vv}}}{N}\sum_{\mathbf{k}\mathbf{k'}\mathbf{q},\sigma_1\sigma_2, \{a,b,c \}}c^{\dagger}_{\mathbf{k}a\sigma_1}c_{\mathbf{k+q}b\sigma_2}c^{\dagger}_{\mathbf{k'}b\sigma_2}c_{\mathbf{k'-q}a\sigma_1}F^{\text{vv}}_c\left(\mathbf{k},\mathbf{k'},\mathbf{q}\right)\nonumber\\-&\frac{\mathcal{V}_{\text{sv}}}{N}\sum_{\mathbf{k}\mathbf{k'}\mathbf{q},\sigma_1\sigma_2,a}
c^{\dagger}_{\mathbf{k}s\sigma_1}c_{\mathbf{k+q}a\sigma_2}c^{\dagger}_{\mathbf{k'}a\sigma_2}c_{\mathbf{k'-q}s\sigma_1} F^{\text{sv}}_a\left(\mathbf{k},\mathbf{k'},\mathbf{q}\right)\label{eq:NN_Coulomb_interacting_ham_k}
\end{align}
In the above, the sum over the indices $(a,b,c)$ are cyclic permutations of $(A,B,C)$ and $a \in \{A,B,C\}$. We have introduced form factors for V-V interaction $F^{\text{vv}}$ as
\begin{gather}
F^{\text{vv}}_c\left(\mathbf{k},\mathbf{k'},\mathbf{q}\right)=\left(1+e^{-i\mathbf{x}_{c}\cdot \left(\mathbf{k+q-k'} \right)}\right) .\label{eq:vv_ff_old}
\end{gather}
Here, $c$ index specifying the bond direction $\hat{\text{w}}_c$ (see Fig.~\ref{fig:1_supp} (a)), and the vector $\mathbf{x}_{c}$ is defined as follows, $(\mathbf{x}_{C}, \mathbf{x}_{A},\mathbf{x}_{B} )=(\mathbf{R}_1, \mathbf{R}_2-\mathbf{R}_1, -\mathbf{R}_2 )$ (also shown in Fig.~\ref{fig:1_supp} (a) and caption). We have also introduced form factors $F^{\text{sv}}$ for interactions between V and Sb-ip as,
\begin{gather}
F^{\text{sv}}_A\left(\mathbf{k},\mathbf{k'},\mathbf{q}\right)=\left(e^{+i\mathbf{x}_C\cdot \left(\mathbf{k}+\mathbf{q}-\mathbf{k'} \right)}+e^{-i\mathbf{x}_B\cdot \left(\mathbf{k}+\mathbf{q}-\mathbf{k'} \right)} \right),\\ F^{\text{sv}}_B\left(\mathbf{k},\mathbf{k'},\mathbf{q}\right)=\left(1+e^{-i\mathbf{x}_B\cdot \left(\mathbf{k}+\mathbf{q}-\mathbf{k'} \right)} \right),\\
F^{\text{sv}}_C\left(\mathbf{k},\mathbf{k'},\mathbf{q}\right)=\left(1+e^{+i\mathbf{x}_C\cdot \left(\mathbf{k}+\mathbf{q}-\mathbf{k'} \right)} \right).
\end{gather}
We can further split the interaction vertex defined in Eq.~\eqref{eq:NN_Coulomb_interacting_ham_k} into two separate pieces; one vertex can be interpreted as an interaction vertex of the loop-current bond order (LCBO) channel (discussed later), and another in a different channel. For the V-V interaction, we can rewrite the V-V form factor $F^{\text{vv}}_c$ defined in Eq.~\eqref{eq:vv_ff_old} as:
\begin{gather}
F^{\text{vv}}_c\left(\mathbf{k},\mathbf{k'},\mathbf{q}\right)=\frac{1}{2}\left[ 
\left(1+ e^{-i\mathbf{x}_{c}\cdot \left(\mathbf{k+q} \right)}\right) \left(1+ e^{+i\mathbf{x}_{c}\cdot \mathbf{k'} } \right)+\left(1- e^{-i\mathbf{x}_{c}\cdot \left(\mathbf{k+q} \right)}\right) \left(1- e^{+i\mathbf{x}_{c}\cdot \mathbf{k'} } \right)\right]\nonumber\\=\frac{1}{2}\left[f_{c,\text{vv}}^{(+)}\left(\mathbf{k+q}\right)f_{c,\text{vv}}^{*(+)}\left(\mathbf{k'}\right)+f_{c,\text{vv}}^{(-)}\left(\mathbf{k+q}\right)f_{c,\text{vv}}^{*(-)}\left(\mathbf{k'}\right)\right]\label{eq:FF_split_vv}.
\end{gather}
Here we have defined form factors for V-V LCBO channel (denoted with $-$) and the other channel (denoted with $+$), 
\begin{gather}
f_{c,\text{vv}}^{(\pm)}\left(\mathbf{k}\right)=1\pm e^{-i \mathbf{x}_c\cdot \mathbf{k}}\label{eq:vv_ff_new}.
\end{gather} 
Using the new form factors defined in Eq.~\eqref{eq:vv_ff_new}, we can rewrite the V-V interaction Hamiltonian from Eq.~\eqref{eq:NN_Coulomb_interacting_ham_k} as
\begin{gather}
H_{\text{int,vv}}=-\frac{\mathcal{V}_{\text{vv}}}{2N}\sum_{\mathbf{q},\sigma_1\sigma_2,a} \mathcal{\hat{L}}^{(+),\text{vv}}_{a,\sigma_1\sigma_2}\left( \mathbf{q}\right) \left[\mathcal{\hat{L}}^{(+),\text{vv}}_{a,\sigma_1\sigma_2}\left( \mathbf{q}\right)\right]^{\dagger}-\frac{\mathcal{V}_{\text{vv}}}{2N}\sum_{\mathbf{q},\sigma_1\sigma_2,a} \mathcal{\hat{L}}^{(-),\text{vv}}_{a,\sigma_1\sigma_2}\left( \mathbf{q}\right) \left[\mathcal{\hat{L}}^{(-),\text{vv}}_{a,\sigma_1\sigma_2}\left( \mathbf{q}\right)\right]^{\dagger}  .
\end{gather}
In the above, we have used the $\mathcal{\hat{L}}$ operator, having three components: $\mathcal{\hat{L}}(\mathbf{q})=\left(\mathcal{\hat{L}}_{A}(\mathbf{q}),\mathcal{\hat{L}}_{B}(\mathbf{q}),\mathcal{\hat{L}}_{C}(\mathbf{q}) \right)$ (other indices suppressed for clarity) along the three bond directions $\hat{\text{w}}$ in Fig.~\ref{fig:1_supp} (a). In general for each component, condensation or ordering may occur for a different momentum $\mathbf{q}=\mathbf{q}_A, \mathbf{q}_B, \mathbf{q}_C$. The component of $\mathcal{\hat{L}}(\mathbf{q})$ along the bond direction $\hat{\text{w}}_c$ is defined as
\begin{gather}
\mathcal{\hat{L}}^{(\pm),\text{vv}}_{c,\sigma_1\sigma_2}\left( \mathbf{q}\right)=\sum_{\mathbf{k}}\Tilde{\varepsilon}_{abc}c^{\dagger}_{\mathbf{k}a\sigma_1}c_{\mathbf{k+q}b\sigma_2}f_{c,\text{vv}}^{\pm}\left(\mathbf{k+q}\right) 
\label{eq:define_L_+-},
\end{gather}
where we use the $\Tilde{\varepsilon}_{abc}$ tensor defined as, $\Tilde{\varepsilon}_{ABC}=\Tilde{\varepsilon}_{BCA}=\Tilde{\varepsilon}_{CAB}=1$, else $\Tilde{\varepsilon}_{abc}=0$. 

 The coexisting $2\times 2$ loop current bond order (LCBO) on the V bonds of the kagome lattice is characterized by a 3-component order parameter, where each component specifies the order along a specific bond direction $\hat{\text{w}}$ (see Fig.~\ref{fig:1_supp} (a)). More details can be found in \cite{Denner_2021,Li_2023,Li_2024}. At $\mathbf{q}=M_{C}$, we see from Eq.~\eqref{eq:define_L_+-} that $\mathcal{\hat{L}}^{(-),\text{vv}}_{C,\sigma_1\sigma_2}(M_{C})$ is the component of this $2\times 2$ LCBO order parameter along the $\hat{\text{w}}_C$ direction. This correspondence can be easily identified when the order parameter is written in real space \cite{Denner_2021,Li_2023,Li_2024},
\begin{gather}
\mathcal{\hat{L}}^{(-),\text{vv}}_{C,\sigma_1\sigma_2}(M_{C})\sim \sum_{\mathbf{r}_j}\left( c^{\dagger}_{\mathbf{r}_jA\sigma_1}c_{\mathbf{r}_jB\sigma_2}-c^{\dagger}_{\mathbf{r}_jA\sigma_1}c_{(\mathbf{r}_j-\mathbf{x}_c)B\sigma_2}\right) \cos(M_C\cdot \mathbf{r}_j).
\end{gather}
The full three-component order parameter for the $2\times 2$ LCBO order on the V bonds on the kagome lattice is 
\begin{gather}
\mathcal{\hat{L}}^{\text{vv}}_{2\times 2}=\left(\mathcal{\hat{L}}^{(-),\text{vv}}_{A}\left(\mathbf{q}_A=M_A\right),\mathcal{\hat{L}}^{(-),\text{vv}}_{B}\left( \mathbf{q}_B=M_B\right),\mathcal{\hat{L}}^{(-),\text{vv}}_{C}\left(\mathbf{q}_C=M_C\right)\right)
\end{gather}
In the condensed state such $2\times 2$ orders have three ordering wave vectors, $\mathbf{q}_A=M_A,\mathbf{q}_B=M_B,\mathbf{q}_C=M_C$, and are also called $3\mathbf{Q}$ states. The pattern of such V-V bond-orders or V-V loop-currents in the condensed state is shown in Figs.~\ref{fig:3_supp} (a,b).

However in the $(+)$ channel, $\mathcal{\hat{L}}^{(+),\text{vv}}_{a,\sigma_1\sigma_2}(M_{a})$ is not an order parameter for any experimentally observed ordering. Since we are interested in computing the fluctuations of the loop-current bond-order phases, henceforth we do not consider fluctuations in the $\mathcal{L}^{(+),\text{vv}}$ channel. This is justified as we eventually expect the $\mathcal{\hat{L}}^{(-),\text{vv}}_{a}$ channel to condense \cite{Li_2023,Li_2024}, and therefore the $\mathcal{\hat{L}}^{(-),\text{vv}}_{a}$ channel would have enhanced fluctuations. Indeed, by computing the RPA-corrected susceptibilities (discussed later), we find that fluctuations in the $\mathcal{\hat{L}}^{(+)}$ channel are smaller than the fluctuations in the LCBO $\mathcal{\hat{L}}^{(-)}$ channel, further justifying this approximation. 

Similar to V-V form-factor in Eq.~\eqref{eq:FF_split_vv}, we can similarly split the V-Sb-ip form factor $F^{\text{sv}}_a\left(\mathbf{k},\mathbf{k'},\mathbf{q}\right)$. For the $\hat{\text{w}}_C$ bond direction,
\begin{gather}
F^{\text{sv}}_C\left(\mathbf{k},\mathbf{k'},\mathbf{q}\right)
=\frac{1}{2}\left[f_{C,\text{sv}}^{(+)}\left(\mathbf{k+q}\right)f_{C,\text{sv}}^{*(+)}\left(\mathbf{k'}\right)+f_{C,\text{sv}}^{(-)}\left(\mathbf{k+q}\right)f_{C,\text{sv}}^{*(-)}\left(\mathbf{k'}\right) \right].
\end{gather}
In the above, we have defined new form factors for V-Sb-ip channels similar to that in Eq.~\eqref{eq:vv_ff_new},
\begin{gather}
f_{A,\text{sv}}^{(\pm)}\left(\mathbf{k}\right)= e^{-i \mathbf{x}_B\cdot \mathbf{k}}\pm e^{+i \mathbf{x}_C\cdot \mathbf{k}},\quad f_{B,\text{sv}}^{(\pm)}\left(\mathbf{k}\right)= e^{-i \mathbf{x}_B\cdot \mathbf{k}}\pm 1,\quad f_{C,\text{sv}}^{(\pm)}\left(\mathbf{k}\right)=1\pm e^{+i \mathbf{x}_C\cdot \mathbf{k}}.
\end{gather}
Similar to Eq.~\eqref{eq:define_L_+-} we defined,
\begin{gather}
\mathcal{\hat{L}}^{(\pm),\text{sv}}_{a,\sigma_1\sigma_2}\left( \mathbf{q}\right)=\sum_{\mathbf{k}}c^{\dagger}_{\mathbf{k}s\sigma_1}c_{\mathbf{k+q}a\sigma_2}f_{a,\text{sv}}^{(\pm)}\left(\mathbf{k+q}\right)\label{eq:eq:define_L_+-_sv}.
\end{gather}
Condensation of $\left(\mathcal{\hat{L}}^{(-),\text{sv}}_{A}\left(M_{A}\right),\mathcal{\hat{L}}^{(-),\text{sv}}_{B}\left( M_{B}\right),\mathcal{\hat{L}}^{(-),\text{sv}}_{C}\left( M_{C}\right)\right)$, would correspond to a $2\times 2$ coexisting LCBO order between the V and Sb-ip sites. The condensed patterns of such $3\mathbf{Q}$ ordered V-Sb-ip bond-order and loop-current are shown separately in Figs.~\ref{fig:3_supp} (c,d). However, V-Sb-ip bond-orders or V-Sb-ip loop-currents have not been detected in experiments. However, this does not rule out the presence of fluctuations of these channels in the disordered state. Therefore, to gain more insight into the role of different fluctuating bond-order/loop-current channels in mediating superconductivity, we consider the effects of fluctuations of both the V-V and V-Sb-ip channels. Also, similar to the V-V channel, we do not consider the effects of the $\mathcal{\hat{L}}^{(+),\text{sv}}$ fluctuating channel. 

Therefore, in the interaction Hamiltonian we henceforth do not consider the ($+$) channel and only consider the LCBO/($-$) channel vertices on the V-V and V-Sb-ip bonds. We also rescale the interaction $\mathcal{V}_{\text{vv/sv}}\rightarrow 2\mathcal{V}_{\text{vv/sv}}$, so that the Hamiltonian has the form as below
\begin{gather}
H_{\text{int}}=-\frac{\mathcal{V}_{\text{vv}}}{N}\sum_{\mathbf{q},\sigma_1\sigma_2,a} \mathcal{\hat{L}}^{(-),\text{vv}}_{a,\sigma_1\sigma_2}\left( \mathbf{q}\right) \left[\mathcal{\hat{L}}^{(-),\text{vv}}_{a,\sigma_1\sigma_2}\left( \mathbf{q}\right)\right]^{\dagger}-\frac{\mathcal{V}_{\text{sv}}}{N}\sum_{\mathbf{q},\sigma_1\sigma_2,a}\mathcal{\hat{L}}^{(-),\text{sv}}_{a,\sigma_1\sigma_2}\left( \mathbf{q}\right) \left[\mathcal{\hat{L}}^{(-),\text{sv}}_{a,\sigma_1\sigma_2}\left( \mathbf{q}\right)\right]^{\dagger}\label{eq:final_interaction_ham}.
\end{gather}
Henceforth, for ease of notation, we will drop the $(-)$ superscript, $\mathcal{\hat{L}}^{(-)}\rightarrow \mathcal{\hat{L}}$ when referring to the LCBO channels. Similarly, for all the form-factors henceforth we use $f^{(-)}\rightarrow f$.

\subsection{Bond orders and Loop current vertices in the chare and spin channels}
The bare-vertex in Eq.~\eqref{eq:final_interaction_ham}, involves interactions between the LCBO vertices with the spin components as specified. However, it can be re-written in the charge and spin channels. Using the identity, $\delta_{\alpha \delta}\delta_{\beta \gamma}=\frac{1}{2}\delta_{\alpha \beta} \delta_{\gamma \delta}+\frac{1}{2}\Vec{\sigma}_{
\alpha \beta} \cdot \Vec{\sigma}_{
\gamma \delta}$, Eq.~\eqref{eq:final_interaction_ham} can be written as,
\begin{align}
H_{\text{int}}=&-\frac{\mathcal{V}_{\text{vv}}}{2N}\sum_{\mathbf{q},a} \mathcal{\hat{L}}^{c,\text{vv}}_{a}\left( \mathbf{q}\right) \left[\mathcal{\hat{L}}^{c,\text{vv}}_{a}\left( \mathbf{q}\right)\right]^{\dagger}
-\frac{\mathcal{V}_{\text{vv}}}{2N}\sum_{\mathbf{q},a} \mathcal{\hat{L}}^{s,\text{vv}}_{\textit{a}}\left( \mathbf{q}\right) \cdot\left[\mathcal{\hat{L}}^{s,\text{vv}}_{\textit{a}}\left( \mathbf{q}\right)\right]^{\dagger}\nonumber\\
&-\frac{\mathcal{V}_{\text{sv}}}{2N}\sum_{\mathbf{q},\textit{a}}\mathcal{\hat{L}}^{c,\text{sv}}_{a}\left( \mathbf{q}\right) \left[\mathcal{\hat{L}}^{c,\text{sv}}_{a}\left( \mathbf{q}\right)\right]^{\dagger}-\frac{\mathcal{V}_{\text{sv}}}{2N}\sum_{\mathbf{q},\textit{a}}\mathcal{\hat{L}}^{s,\text{sv}}_{\textit{a}}\left( \mathbf{q}\right) \cdot\left[\mathcal{\hat{L}}^{s,\text{sv}}_{\textit{a}}\left( \mathbf{q}\right)\right]^{\dagger}\label{eq:final_interaction_ham_1}.
\end{align}
In the above Eq.~\eqref{eq:final_interaction_ham_1}, the LCBO order parameter has been defined in the charge ($\texttt{i}=c$) and spin ($\texttt{i}=s$) channels using Eqs.~\eqref{eq:define_L_+-},~\eqref{eq:eq:define_L_+-_sv} as: $\mathcal{\hat{L}}^{\texttt{i}}_{a}=\sum_{\sigma_1\sigma_2}\Gamma^{\texttt{i}}_{\sigma_1\sigma_2}\mathcal{\hat{L}}_{a,\sigma_1\sigma_2}$, where $\Gamma^{\texttt{i}}=(\Gamma^c,\Gamma^s)=(\mathds{1},\Vec{\boldsymbol{\sigma}})$, where we have suppressed the other indices for simplicity. 

We can split the LCBO vertex into separate vertices between bond-orders (BO) and loop-currents (LC) by taking its real and imaginary parts respectively. This splitting is performed for both the V-V and V-Sb-ip interactions, and also separately for the charge and spin channels. This introduces eight different order parameters, with each order parameter defined on the V-V (vv) or V-Sb-ip (sv) bonds: \textit{charge bond-orders}: $\mathcal{\hat{B}}^{c,\text{vv}}_{ \mathbf{q}}$, $\mathcal{\hat{B}}^{c,\text{sv}}_{ \mathbf{q}}$; \textit{charge loop-currents}: $\mathcal{\hat{J}}^{c,\text{vv}}_{ \mathbf{q}}$, $\mathcal{\hat{J}}^{c,\text{sv}}_{ \mathbf{q}}$; \textit{spin bond-orders}: $\mathcal{\hat{B}}^{s,\text{vv}}_{ \mathbf{q}}$, $\mathcal{\hat{B}}^{s,\text{sv}}_{ \mathbf{q}}$; \textit{spin loop-currents} $\mathcal{\hat{J}}^{s,\text{vv}}_{ \mathbf{q}}$, $\mathcal{\hat{J}}^{s,\text{sv}}_{ \mathbf{q}}$. Further, here each order parameter is made up of three components, \textit{e.g.} $\mathcal{\hat{B}}^{\textit{i},\text{vv}}_{\mathbf{q}}=\left(\mathcal{\hat{B}}^{\textit{i},\text{vv}}_{\mathbf{q},A},\mathcal{\hat{B}}^{\textit{i},\text{vv}}_{\mathbf{q},B},\mathcal{\hat{B}}^{\textit{i},\text{vv}}_{\mathbf{q},C}\right)$, specifying the orders along the three bond directions $\hat{\text{w}}$ shown in Fig.~\ref{fig:1_supp} (a). Condensation or ordering can occur at different values of $\mathbf{q}=\mathbf{q}_A,\mathbf{q}_B,\mathbf{q}_C$ for each of the three components. The different order parameters, along the bond-direction $\hat{\text{w}}_a$ (see Fig.~\ref{fig:1_supp} (a)), are defined as follows,
\begin{gather}
\mathcal{\hat{B}}^{\texttt{i},\text{vv}}_{ \mathbf{q},\textit{a}}=\frac{1}{2}\left( \mathcal{\hat{L}}^{\texttt{i},\text{vv}}_{ \mathbf{q},\textit{a}}+\left[\mathcal{\hat{L}}^{\texttt{i},\text{vv}}_{ \mathbf{-q},\textit{a}}\right]^{\dagger}\right)=\sum_{\mathbf{k},\sigma_1\sigma_2}\frac{\Tilde{\varepsilon}_{abc}}{2}\left(c^{\dagger}_{\mathbf{k},b,\sigma_1}c_{\mathbf{k+q},c,\sigma_2}f_{a,\text{vv}}\left(\mathbf{k+q}\right)+ c^{\dagger}_{\mathbf{k},c,\sigma_1}c_{\mathbf{k+q},b,\sigma_2}f_{a,\text{vv}}^{*}\left(\mathbf{k}\right)\right)\Gamma^{\texttt{i}}_{\sigma_1\sigma_2},
\end{gather}
\begin{gather}
\mathcal{\hat{J}}^{\texttt{i},\text{vv}}_{ \mathbf{q},\textit{a}}=\frac{i}{2}\left(-\mathcal{\hat{L}}^{\texttt{i},\text{vv}}_{ \mathbf{q},\textit{a}}+\left[\mathcal{\hat{L}}^{\texttt{i},\text{vv}}_{ \mathbf{-q},\textit{a}}\right]^{\dagger} \right)=\sum_{\mathbf{k},\sigma_1\sigma_2}\frac{-i\Tilde{\varepsilon}_{abc}}{2}\left(c^{\dagger}_{\mathbf{k},b,\sigma_1}c_{\mathbf{k+q},c,\sigma_2}f_{a,\text{vv}}\left(\mathbf{k+q}\right)-c^{\dagger}_{\mathbf{k},c,\sigma_1}c_{\mathbf{k+q},b,\sigma_2}f_{a,\text{vv}}^{*}\left(\mathbf{k}\right)\right)\Gamma^{\texttt{i}}_{\sigma_1\sigma_2},
\end{gather}
\begin{gather}
\mathcal{\hat{B}}^{\texttt{i},\text{sv}}_{ \mathbf{q},\textit{a}}=\frac{1}{2}\left( \mathcal{\hat{L}}^{\texttt{i},\text{sv}}_{ \mathbf{q},\textit{a}}+\left[\mathcal{\hat{L}}^{\texttt{i},\text{sv}}_{ \mathbf{-q},\textit{a}}\right]^{\dagger}\right)=\sum_{\mathbf{k},\sigma_1\sigma_2}\frac{\Tilde{\varepsilon}_{abc}}{2}\left(c^{\dagger}_{\mathbf{k},s,\sigma_1}c_{\mathbf{k+q},a,\sigma_2}f_{a,\text{sv}}\left(\mathbf{k+q}\right)+ c^{\dagger}_{\mathbf{k},a,\sigma_1}c_{\mathbf{k+q},s,\sigma_2}f_{a,\text{sv}}^{*}\left(\mathbf{k}\right)\right)\Gamma^{\texttt{i}}_{\sigma_1\sigma_2},
\end{gather}
\begin{gather}
\mathcal{\hat{J}}^{\texttt{i},\text{sv}}_{ \mathbf{q},\textit{a}}=\frac{i}{2}\left(-\mathcal{\hat{L}}^{\texttt{i},\text{vs}}_{ \mathbf{q},\textit{a}}+\left[\mathcal{\hat{L}}^{\texttt{i},\text{sv}}_{ \mathbf{-q},\textit{a}}\right]^{\dagger} \right)=\sum_{\mathbf{k},\sigma_1\sigma_2}\frac{-i\Tilde{\varepsilon}_{abc}}{2}\left(c^{\dagger}_{\mathbf{k},s,\sigma_1}c_{\mathbf{k+q},a,\sigma_2}f_{a,\text{sv}}\left(\mathbf{k+q}\right)-c^{\dagger}_{\mathbf{k},a,\sigma_1}c_{\mathbf{k+q},s,\sigma_2}f_{a,\text{sv}}^{*}\left(\mathbf{k}\right)\right)\Gamma^{\texttt{i}}_{\sigma_1\sigma_2},
\end{gather}
where $(a,b,c)$ are cyclic permutations of $(A,B,C)$, and we have used the tensor $\Tilde{\varepsilon}_{abc}$ defined as: $\Tilde{\varepsilon}_{ABC}=\Tilde{\varepsilon}_{BCA}=\Tilde{\varepsilon}_{CAB}=1$, else $\Tilde{\varepsilon}_{abc}=0$. Also, $\texttt{i}\in \{c,s\}$ denotes the charge (c) and spin (s) channels respectively. The BO ($\mathcal{B}_{\mathbf{q}}$) and LC ($\mathcal{J}_{\mathbf{q}}$) order parameters defined above are hermitian, $\left[\mathcal{\hat{O}}_{\mathbf{q}}\right]^{\dagger}=\mathcal{\hat{O}}_{\mathbf{-q}}$.
The final bare-interaction Hamiltonian Eq.~\eqref{eq:final_interaction_ham_1}, can be explicitly written in all the eight BO and LC channels as follows,
\begin{align}
H_{\text{int}}=-\frac{\mathcal{V}_{\text{vv}}}{2N}\sum_{\mathbf{q},a} \left(\mathcal{\hat{B}}^{c,\text{vv}}_{\mathbf{q},a}\mathcal{\hat{B}}^{c,\text{vv}}_{\mathbf{-q},a}+\mathcal{\hat{B}}^{s,\text{vv}}_{\mathbf{q},a}\cdot\mathcal{\hat{B}}^{s,\text{vv}}_{\mathbf{-q},a}
+
\mathcal{\hat{J}}^{c,\text{vv}}_{\mathbf{q},a}\mathcal{\hat{J}}^{c,\text{vv}}_{\mathbf{-q},a}+\mathcal{\hat{J}}^{s,\text{vv}}_{\mathbf{q},a}\cdot\mathcal{\hat{J}}^{s,\text{vv}}_{\mathbf{-q},a}
\right)\nonumber\\
-\frac{\mathcal{V}_{\text{sv}}}{2N}\sum_{\mathbf{q},a}\left(
\mathcal{\hat{B}}^{c,\text{sv}}_{\mathbf{q},a}\mathcal{\hat{B}}^{c,\text{sv}}_{\mathbf{-q},a}+\mathcal{\hat{B}}^{s,\text{sv}}_{\mathbf{q},a}\cdot\mathcal{\hat{B}}^{s,\text{sv}}_{\mathbf{-q},a}
+
\mathcal{\hat{J}}^{c,\text{sv}}_{\mathbf{q},a}\mathcal{\hat{J}}^{c,\text{sv}}_{\mathbf{-q},a}+\mathcal{\hat{J}}^{s,\text{sv}}_{\mathbf{q},a}\cdot\mathcal{\hat{J}}^{s,\text{sv}}_{\mathbf{-q},a}
\right)\label{eq:final_interaction_ham_split},
\end{align}
where $a\in\{A,B,C\}$, and it specifies the different bond directions for the three components of the order parameter. In the main text, Eq.~\eqref{eq:final_interaction_ham_split} is expressed in a compact form. 

Additionally, we find that the BO and LC order parameters in the charge and spin channels have definite parities under time-reversal $\mathcal{T}$. Charge bond orders and spin loop currents are time-reversal even with $P_T=+1$, and spin bond orders and charge loop currents are time-reversal odd  with $P_T=-1$.
\begin{gather}
\mathcal{T}\mathcal{\hat{B}}^{c,\text{vv}}_{\mathbf{q},a}\mathcal{T}^{-1}=+1 \mathcal{\hat{B}}^{c,\text{vv}}_{\mathbf{-q},a},\quad\mathcal{T}\mathcal{\hat{B}}^{c,\text{sv}}_{\mathbf{q},a}\mathcal{T}^{-1}=+1 \mathcal{\hat{B}}^{c,\text{sv}}_{\mathbf{-q},a},\\
\mathcal{T}\mathcal{\hat{B}}^{s,\text{vv}}_{\mathbf{q},a}\mathcal{T}^{-1}=-1 \mathcal{\hat{B}}^{s,\text{vv}}_{\mathbf{-q},a},\quad\mathcal{T}\mathcal{\hat{B}}^{s,\text{sv}}_{\mathbf{q},a}\mathcal{T}^{-1}=-1 \mathcal{\hat{B}}^{s,\text{sv}}_{\mathbf{-q},a},\\
\mathcal{T}\mathcal{\hat{J}}^{c,\text{vv}}_{\mathbf{q},a}\mathcal{T}^{-1}=-1 \mathcal{\hat{J}}^{c,\text{vv}}_{\mathbf{-q},a},\quad\mathcal{T}\mathcal{\hat{J}}^{c,\text{sv}}_{\mathbf{q},a}\mathcal{T}^{-1}=-1 \mathcal{\hat{J}}^{c,\text{sv}}_{\mathbf{-q},a},\\
\mathcal{T}\mathcal{\hat{J}}^{s,\text{vv}}_{\mathbf{q},a}\mathcal{T}^{-1}=+1 \mathcal{\hat{J}}^{s,\text{vv}}_{\mathbf{-q},a},\quad\mathcal{T}\mathcal{\hat{J}}^{s,\text{sv}}_{\mathbf{q},a}\mathcal{T}^{-1}=+1 \mathcal{\hat{J}}^{s,\text{sv}}_{\mathbf{-q},a}.
\end{gather}
These results will be used in Sec. \ref{sec:T_parity_supp}.

\section{Calculating the Cooper vertex}\label{sec:vertex_Cooper_supp}
\subsection{The bare vertex}
From Eq.~\eqref{eq:final_interaction_ham_split}, the bare interaction vertex in the BO $(\mathcal{B}_{\mathbf{q}}\mathcal{B}_{\mathbf{-q}})$ or LC $(\mathcal{J}_{\mathbf{q}}\mathcal{J}_{\mathbf{-q}})$ channels can be written as components of a tensor with orbital \& spin components. The Hamiltonian is defined in terms of this tensor with orbital $(\mu_i)$ and spin $(\sigma_i)$ components as follows,
\begin{gather}
H=\frac{1}{N}\sum_{\mathbf{k,k',q}}\left[W(\mathbf{k,k',q})\right]^{\mu_1\sigma_1,\mu_2\sigma_2}_{\mu_3\sigma_3,\mu_4\sigma_4}c_{\mathbf{k},\mu_1, \sigma_1}^{\dagger} c_{\mathbf{k'},\mu_3 ,\sigma_3}^{\dagger}
c_{\mathbf{k'-q},\mu_2, \sigma_2} c_{\mathbf{k+q},\mu_4, \sigma_4}.
\end{gather}
Due to the absence of spin-orbit coupling, the orbital sector of the vertex is decoupled from the spin sector. Therefore, we can set the spin-sector of the vertex as follows, where the parameter $\alpha\in[0,1]$ tunes between the charge and spin channels,
\begin{gather}
\Gamma(\alpha)^{\sigma_1 \sigma_4}_{\sigma_3 \sigma_2}=(1-\alpha)\mathds{1}_{\sigma_1 \sigma_4}
\mathds{1}_{\sigma_3 \sigma_2}+\alpha \Vec{\boldsymbol{\sigma}}_{\sigma_1 \sigma_4} \cdot \Vec{\boldsymbol{\sigma}}_{\sigma_3 \sigma_2}.\label{eq:spin_charge_channels}
\end{gather}
This is done to gain more insight into the effects of fluctuations present across both the charge and spin sectors. Using $\Tilde{\mu}=(\mu,\sigma)$ as a combined orbital and spin index we can express the bare vertex in Eq.~\eqref{eq:final_interaction_ham_split} in a symmetric form as below. Henceforth, to denote the V-$\Tilde{d}_{\alpha}$ orbital we use the tensor index $\mu=\alpha$, and for the Sb-ip-$p_z$ orbital we use tensor index $\mu=s$.
\begin{align}
\left[W(\mathbf{k,k',q})\right]^{\Tilde{\mu}_1,\Tilde{\mu}_2}_{\Tilde{\mu}_3,\Tilde{\mu}_4}=&-\frac{\mathcal{V}_{\text{vv}}}{16}
\delta_{\mu_1\mu_2}\delta_{\mu_3\mu_4}
&\left[
\Tilde{\varepsilon}_{\mu_1\mu_4 \alpha}\left(f_{\alpha\text{,vv}}\left(\mathbf{k+q} \right) f_{\alpha\text{,vv}}^*\left(\mathbf{k'} \right) +
f_{\alpha\text{,vv}}\left(\mathbf{k} \right) f_{\alpha\text{,vv}}^*\left(\mathbf{k'-q} \right)
\right)\right.&\nonumber\\
&  &+\left.\Tilde{\varepsilon}_{\mu_4\mu_1 \alpha}
\left(
f_{\alpha\text{,vv}}^*\left(\mathbf{k+q} \right) f_{\alpha\text{,vv}}\left(\mathbf{k'} \right) +
f_{\alpha\text{,vv}}^*\left(\mathbf{k} \right) f_{\alpha\text{,vv}}\left(\mathbf{k'-q} \right)\right)
\right]&\Gamma(\alpha)^{\sigma_1 \sigma_4}_{\sigma_3 \sigma_2} \nonumber\\
&\mp\frac{\mathcal{V}_{\text{vv}}}{16}
\delta_{\mu_1\mu_3}\delta_{\mu_2\mu_4}
&
\left[
\Tilde{\varepsilon}_{\mu_1\mu_4 \alpha}\left(f_{\alpha\text{,vv}}\left(\mathbf{k+q} \right) f_{\alpha\text{,vv}}\left(\mathbf{k'-q} \right) +
f_{\alpha\text{,vv}}\left(\mathbf{k} \right) f_{\alpha\text{,vv}}\left(\mathbf{k'} \right)
\right)\right.&\nonumber\\
&  &+\left.\Tilde{\varepsilon}_{\mu_4\mu_1 \alpha}
\left(
f_{\alpha\text{,vv}}^*\left(\mathbf{k+q} \right) f_{\alpha\text{,vv}}^*\left(\mathbf{k'-q} \right) +
f_{\alpha\text{,vv}}^*\left(\mathbf{k} \right) f_{\alpha\text{,vv}}^*\left(\mathbf{k'} \right)\right)
\right]&\Gamma(\alpha)^{\sigma_1 \sigma_4}_{\sigma_3 \sigma_2} \nonumber\\
& -\frac{\mathcal{V}_{\text{vs}}}{16} \delta_{\mu_1\mu_2}\delta_{\mu_3\mu_4}\xi_{\mu_1\mu_3} & \left[\delta_{\mu_1s}\left(f_{\mu_3\text{,vs}}
\left(\mathbf{k+q} \right) f_{\mu_3\text{,vs}}^*\left(\mathbf{k'} \right) +
f_{\mu_3\text{,vs}}\left(\mathbf{k} \right) f_{\mu_3\text{,vs}}^*\left(\mathbf{k'-q} \right) \right)\right. & \nonumber\\
&  & +\left.\delta_{\mu_3s}
\left(f_{\mu_1\text{,vs}}^*
\left(\mathbf{k+q} \right) f_{\mu_1\text{,vs}}\left(\mathbf{k'} \right) +
f_{\mu_1\text{,vs}}^*\left(\mathbf{k} \right) f_{\mu_1\text{,vs}}\left(\mathbf{k'-q} \right) \right)\right] & \Gamma(\alpha)^{\sigma_1 \sigma_4}_{\sigma_3 \sigma_2} \nonumber\\
& \mp \frac{\mathcal{V}_{\text{vs}}}{16} \delta_{\mu_1\mu_3}\delta_{\mu_2\mu_4}\xi_{\mu_1\mu_2} & \left[\delta_{\mu_1s}\left(f_{\mu_2\text{,vs}}
\left(\mathbf{k+q} \right) f_{\mu_2\text{,vs}}\left(\mathbf{k'-q} \right) +
f_{\mu_2\text{,vs}}\left(\mathbf{k} \right) f_{\mu_2\text{,vs}}\left(\mathbf{k'} \right) \right)\right. & \nonumber\\
&  & +\left.\delta_{\mu_2s}
\left(f_{\mu_1\text{,vs}}^*
\left(\mathbf{k+q} \right) f_{\mu_1\text{,vs}}^*\left(\mathbf{k'-q} \right) +
f_{\mu_1\text{,vs}}^*\left(\mathbf{k} \right) f_{\mu_1\text{,vs}}^*\left(\mathbf{k'} \right) \right)\right] & \Gamma(\alpha)^{\sigma_1 \sigma_4}_{\sigma_3 \sigma_2} \label{eq:full_bare_vertex}.
\end{align}
Here the $+$ is for the BO vertex $(\mathcal{B}_{\mathbf{q}}\mathcal{B}_{\mathbf{-q}})$, and $-$ is for the LC vertex $(\mathcal{J}_{\mathbf{q}}\mathcal{J}_{\mathbf{-q}})$. In Eq.~\eqref{eq:full_bare_vertex}, we have used the $\Tilde{\varepsilon}_{\mu_1\mu_2\mu_3}$ tensor, defined as $\Tilde{\varepsilon}_{ABC}=\Tilde{\varepsilon}_{BCA}=\Tilde{\varepsilon}_{CAB}=1$, else $\Tilde{\varepsilon}_{\mu_1\mu_2\mu_3}=0$. We have also defined another tensor as $\xi_{\mu_1\mu_2}=1$ if $\{ \mu_1\neq \mu_2$ and $\{\mu_1\mu_2\} \in \{A,B,C,s\} \}$, else $\xi_{\mu_1\mu_2}=0$.
In the Cooper channel, the bare vertex can be expressed as
$
\sum_{\mathbf{k,k'}}\left[V_0(\mathbf{k,k'})\right]^{\Tilde{\mu}_1,\Tilde{\mu}_2}_{\Tilde{\mu}_3,\Tilde{\mu}_4}c_{\mathbf{k}\Tilde{\mu}_1}^{\dagger} c_{\mathbf{-k}\Tilde{\mu}_3}^{\dagger}
c_{\mathbf{-k'}\Tilde{\mu}_2} c_{\mathbf{k'}\Tilde{\mu}_4}
$. Eq.~\eqref{eq:full_bare_vertex}, can be written in the Cooper channel as follows: $[V_0(\mathbf{k},\mathbf{k'})]=[W(\mathbf{k},-\mathbf{k},\mathbf{k'}-\mathbf{k})]$.

\subsection{Generalized bare-susceptibility}
We define a generalized bare-susceptibility $[\chi_0]$ as below. Here $[\chi_0]$ appears as a repeating structure in the RPA-bubble re-summation using the BO/LC vertices. 
\begin{align}
\left[\chi_0(\mathbf{q},0)\right]^{\Tilde{\nu}_1,\Tilde{\nu}_2}_{\Tilde{\nu}_3,\Tilde{\nu}_4} &=
-\frac{1}{\beta N}\sum_{\mathbf{p},ip_n}\mathcal{G}_{0,\Tilde{\nu}_2, \Tilde{\nu}_3}(\mathbf{p}+\mathbf{q},ip_n+0)\mathcal{G}_{0,\Tilde{\nu}_4, \Tilde{\nu}_1}(\mathbf{p},ip_n)\left[f_{\chi}(\mathbf{p},\mathbf{q})\right]^{\nu_1,\nu_2}_{\nu_3,\nu_4}\\
&=-\frac{1}{N}\sum_{\mathbf{p},n_1n_2}\left[\Tilde{M}(\mathbf{p},\mathbf{q})\right]_{n_1n_2}^{\Tilde{\nu}_1,\Tilde{\nu}_2,\Tilde{\nu}_3,\Tilde{\nu}_4}\frac{n_F(\xi_{\mathbf{p},n_1})-n_F(\xi_{\mathbf{p+q},n_2})}{0+\xi_{\mathbf{p},n_1}-\xi_{\mathbf{p+q},n_2}}\left[f_{\chi}(\mathbf{p},\mathbf{q})\right]^{\nu_1,\nu_2}_{\nu_3,\nu_4}\label{eq:bare_susc}
\end{align}
where we use the following definitions
\begin{align}
&\left[f_{\chi}(\mathbf{p},\mathbf{q})\right]^{\nu_1,\nu_2}_{\nu_3,\nu_4}=\left[f_{\chi,\text{vv}}(\mathbf{p},\mathbf{q})\right]^{\nu_1,\nu_2}_{\nu_3,\nu_4}+\left[f_{\chi,\text{sv}}(\mathbf{p},\mathbf{q})\right]^{\nu_1,\nu_2}_{\nu_3,\nu_4}\\
&\left[f_{\chi,\text{vv}}(\mathbf{p},\mathbf{q})\right]^{\nu_1,\nu_2}_{\nu_3,\nu_4}=
\begin{pmatrix}
\sum_{\lambda}\Tilde{\varepsilon}_{\lambda \nu_1 \nu_2}f_{\lambda,\text{vv}}(\mathbf{p}+\mathbf{q})+\Tilde{\varepsilon}_{\lambda \nu_2 \nu_1} f^{*}_{\lambda,\text{vv}}(\mathbf{p})\\
\sum_{\lambda}\Tilde{\varepsilon}_{\lambda \nu_1 \nu_2}f_{\lambda,\text{vv}}(\mathbf{p})+\Tilde{\varepsilon}_{\lambda \nu_2 \nu_1} f^{*}_{\lambda,\text{vv}}(\mathbf{p}+\mathbf{q})
\end{pmatrix}_{2\times 1}
\cdot
\begin{pmatrix}
  \big(f_{\gamma,\text{vv}}(\mathbf{p})\Tilde{\varepsilon}_{\gamma \nu_3 \nu_4}+ & \big(f_{\gamma,\text{vv}}(\mathbf{p}+\mathbf{q})\Tilde{\varepsilon}_{\gamma \nu_3 \nu_4} \\  
  f^{*}_{\gamma,\text{vv}}(\mathbf{p}+\mathbf{q})\Tilde{\varepsilon}_{\gamma \nu_4 \nu_3}\big) & +f^{*}_{\gamma,\text{vv}}(\mathbf{p})\Tilde{\varepsilon}_{\gamma \nu_4 \nu_3}\big)
\end{pmatrix}_{1\times 2} \\
&\left[f_{\chi,\text{sv}}(\mathbf{p},\mathbf{q})\right]^{\nu_1,\nu_2}_{\nu_3,\nu_4}=\xi_{\nu_1 \nu_2} \xi_{\nu_3 \nu_4}
    \left(\delta_{\nu_1 s}f_{\nu_2,\text{sv}}(\mathbf{p}+\mathbf{q})+\delta_{\nu_2 s} f^{*}_{\nu_1,\text{sv}}(\mathbf{p})
    \right)\left(f_{\nu_4,\text{sv}}(\mathbf{p})\delta_{\nu_3 s} +f^{*}_{\nu_3,\text{sv}}(\mathbf{p}+\mathbf{q})\delta_{\nu_4 s}
    \right)
    \begin{pmatrix}
        1 & 0\\
        0 & 1
    \end{pmatrix}_{2\times 2}
    \\
    &\left[\Tilde{M}(\mathbf{p},\mathbf{q})\right]_{n_1n_2}^{\Tilde{\nu}_1,\Tilde{\nu}_2,\Tilde{\nu}_3,\Tilde{\nu}_4}=\left[U(\mathbf{p+q}) \right]_{\mu_2 n_2} \left[U^*(\mathbf{p+q}) \right]_{\mu_3 n_2} \left[U(\mathbf{p})\right]_{\mu_4 n_1} \left[U^*(\mathbf{p}) \right]_{\mu_1 n_1}\delta_{\sigma_2\sigma_3}\delta_{\sigma_4\sigma_1}.
\end{align}
This generalized susceptibility is a $2\times 2$ matrix of the tensors $[\chi]^{\Tilde{\nu_1}\Tilde{\nu_2}}_{\Tilde{\nu_3}\Tilde{\nu_4}}$. This generalized susceptibility is used in subsequent computation of the RPA corrected vertex.
\subsection{RPA-corrected Cooper vertex}
The RPA-corrected Cooper vertex calculated by re-summing bubble diagrams in the particle-hole channel can be written as,
\begin{gather}
\left[V(\mathbf{k},\mathbf{k'})\right]^{\Tilde{\mu}_1\Tilde{\mu}_2}_{\Tilde{\mu}_3\Tilde{\mu}_4}=\left[V_{\text{0}}(\mathbf{k},\mathbf{k'})\right]^{\Tilde{\mu}_1\Tilde{\mu}_2}_{\Tilde{\mu}_3\Tilde{\mu}_4}-\left[V_L(\mathbf{k},\mathbf{k'})\right]^{\Tilde{\mu}_1\Tilde{\nu}_2}_{\Tilde{\nu}_1\Tilde{\mu}_4}\left[\chi_{\text{RPA}}(\mathbf{k}-\mathbf{k'},0)\right]^{\Tilde{\nu}_1\Tilde{\nu}_2}_{\Tilde{\nu}_3\Tilde{\nu}_4}\left[ V_R(\mathbf{k},\mathbf{k'})\right]^{\Tilde{\nu}_3\Tilde{\mu}_2}_{\Tilde{\mu}_3\Tilde{\nu}_4}\label{eq:RPA_susc_gen},
\end{gather}
where the generalized RPA-corrected susceptibility $[\chi_{\text{RPA}}]$ is defied in terms of the generalized bare-susceptibility $[\chi_0]$ as follows,
\begin{gather}
    \chi_{\text{RPA}}(\mathbf{q},0)^{\Tilde{\nu}_1 \Tilde{\nu}_2}_{\Tilde{\nu}_3 \Tilde{\nu}_4}=\left[\left((1+\chi_{0}V_C)^{-1}\chi_{0}\right)(\mathbf{q},0)\right]^{\Tilde{\nu}_1 \Tilde{\nu}_2}_{\Tilde{\nu}_3 \Tilde{\nu}_4}.
\end{gather}
In the above Eq.~\eqref{eq:RPA_susc_gen}, we have used the three tensors left $V_L$, right $V_R$, and center $V_C$ vertex as defined below:
\begin{align}
[V_L(\mathbf{k},\mathbf{k'})]^{\Tilde{\mu}_1 \Tilde{\nu}_2}_{\Tilde{\nu}_1 \Tilde{\mu}_4}=&-\frac{\mathcal{V}_{\text{vv}}}{16}\left(\delta_{\mu_1 \nu_2}\delta_{\nu_1 \mu_4}  \pm \delta_{\mu_1 \nu_1}\delta_{\nu_2 \mu_4}
\right)\Gamma(\alpha)^{\sigma_1\sigma_4}_{\sigma'_1\sigma'_2}
\begin{pmatrix}
    \big(\sum_{\alpha} f_{\alpha,\text{vv}}(\mathbf{k'}) \Tilde{\epsilon}_{\mu_1 \mu_4 \alpha} & \big(\sum_{\alpha} f_{\alpha,\text{vv}}(\mathbf{k}) \Tilde{\epsilon}_{\mu_1 \mu_4 \alpha}\\
    + f^{*}_{\alpha,\text{vv}}(\mathbf{k}) \Tilde{\epsilon}_{\mu_4 \mu_1\alpha}\big) & + f^{*}_{\alpha,\text{vv}}(\mathbf{k'}) \Tilde{\epsilon}_{\mu_4 \mu_1\alpha}\big)
\end{pmatrix}_{1\times 2}\nonumber\\
 &-\frac{\mathcal{V}_{\text{sv}}}{16}\xi_{\mu_1 \mu_4}\left(\delta_{\mu_1 \nu_2}\delta_{\nu_1 \mu_4} \pm \delta_{\mu_1 \nu_1}\delta_{\nu_2 \mu_4}
\right)\Gamma(\alpha)^{\sigma_1\sigma_4}_{\sigma'_1\sigma'_2}\left( f_{\mu_4,\text{sv}}(\mathbf{k'}) \delta_{\mu_1 s} + f^{*}_{\mu_1,\text{sv}}(\mathbf{k}) \delta_{\mu_4 s} \right)
 \begin{pmatrix}
     1 & 1
 \end{pmatrix}_{1\times 2}
 \\
[V_R(\mathbf{k},\mathbf{k'})]^{\Tilde{\nu}_3 \Tilde{\mu}_2}_{\Tilde{\mu}_3 \Tilde{\nu}_4}=&-\frac{\mathcal{V}_{\text{vv}}}{16}\left( \delta_{\nu_3 \mu_2}\delta_{\mu_3 \nu_4} \pm \delta_{\nu_3 \mu_3}\delta_{\mu_2 \nu_4} \right)\Gamma(\alpha)^{\sigma'_3\sigma'_4}_{\sigma_3\sigma_2}
\begin{pmatrix}
\sum_{\beta} \big(f_{\beta,\text{vv}}(-\mathbf{k'}) \Tilde{\epsilon}_{ \mu_3 \mu_2\beta} + f^{*}_{\beta,\text{vv}}(-\mathbf{k}) \Tilde{\epsilon}_{ \mu_2 \mu_3\beta}\big)   \\
\sum_{\beta} \big(f_{\beta,\text{vv}}(-\mathbf{k}) \Tilde{\epsilon}_{ \mu_3 \mu_2\beta} + f^{*}_{\beta,\text{vv}}(-\mathbf{k'}) \Tilde{\epsilon}_{ \mu_2 \mu_3\beta}\big)
\end{pmatrix}_{2\times 1}
\nonumber\\
 &-\frac{\mathcal{V}_{\text{sv}}}{16} \xi_{\mu_2\mu_3}\left( \delta_{\nu_3 \mu_2}\delta_{\mu_3 \nu_4} \pm \delta_{\nu_3 \mu_3}\delta_{\mu_2 \nu_4} \right)\Gamma(\alpha)^{\sigma'_3\sigma'_4}_{\sigma_3\sigma_2}\left( f_{\mu_2,\text{sv}}(-\mathbf{k'}) \delta_{\mu_3 s} + f^{*}_{\mu_3,\text{sv}}(-\mathbf{k}) \delta_{\mu_2 s} \right)
 \begin{pmatrix}
    1 \\
    1
 \end{pmatrix}_{2\times 1}\\
\left[ V_C\right]^{\Tilde{\mu}_1 \Tilde{\nu}_2}_{\Tilde{\nu}_1 \Tilde{\mu}_4}=&-\frac{(\mathcal{V}_{\text{vv}}\eta_{\mu_1\mu_4}+\mathcal{V}_{\text{sv}}\xi_{\mu_1\mu_4})}{16}\left( \delta_{\mu_1 \nu_2}\delta_{\nu_1 \mu_4} \pm \delta_{\mu_1 \nu_1}\delta_{\nu_2 \mu_4} 
\right)\Gamma(\alpha)^{\sigma_1\sigma_4}_{\sigma'_1\sigma'_2}
\begin{pmatrix}
    1 & 0\\
    0 & 1
\end{pmatrix}_{2\times 2}
\end{align}
In the above we have used $\Tilde{\mu}_i=(\mu_i,\sigma_i)$, $\Tilde{\nu}_i=(\nu_i,\sigma'_i)$. Here we have used another tensor defined as $\eta_{\mu_1\mu_2}=1$ if $\{\mu_1\neq\mu_2$ and $\{\mu_1,\mu_2\}\in\{A,B,C \}\}$, else $\eta_{\mu_1\mu_2}=0$. The $\pm$ sign in the above denotes the BO and LC channels respectively.

To study the competition among the different fluctuating channels, we parameterize the relative strengths of the channels by $\{\alpha,\beta,\gamma\}$ as follows. As defined in Eq.~\eqref{eq:spin_charge_channels}, $\alpha$ tunes between the charge and spin vertices. Additionally, in Eq.~\eqref{eq:final_interaction_ham_split} we set $\mathcal{V}_{\text{vv}}=\xi(1-\gamma)$, and $\mathcal{V}_{\text{sv}}=\xi\gamma$, where $\gamma$ tunes between the relative strengths of the V-V and V-Sbip channels, and $\xi$ is an overall interaction scale, which we set as $\xi=4$. Using these parameterizations, we compute the RPA-corrected BO and LC vertices, $V_{\text{BO}}$ and $V_{\text{LC}}$ respectively. The final effective Cooper vertex, considering both BO and LC fluctuations is then written as (keeping momentum and orbital/spin indices implicit)
\begin{gather}
    V(\alpha,\beta,\gamma,\xi)=(1-\beta)V_{\text{BO}}(\alpha,\gamma,\xi)+\beta V_{\text{LC}}(\alpha,\gamma,\xi)\label{eq:all_channels},
\end{gather} 
where $\beta$ tunes between the BO and LC channels. For simplicity, we did not consider mixing among the BO and LC channels in the RPA. We use Eq.~\eqref{eq:all_channels} to construct the phase diagram in Fig. 3(a) of the main text.

\subsection{Bond order/Loop current susceptibility}
The bare (spin-unresolved) BO \& LC susceptibilities can be calculated from the generalized bare-susceptibility in Eq.~\eqref{eq:bare_susc} as,
\begin{gather}
\left[\left\langle \mathcal{\hat{B}}_{\mathbf{q}} \mathcal{\hat{B}}_{\mathbf{-q}} \right\rangle\right]^{\sigma_1\sigma_2}_{\sigma'_1\sigma'_2}=\frac{1}{2}\sum_{(a,b),(lm)}
Tr\left[
\left[\chi_0\right]^{a\sigma_1,b\sigma_2}_{m\sigma'_1,l\sigma'_2}+\left[\chi_0\right]^{b\sigma_1,a\sigma_2}_{l\sigma'_1,m\sigma'_2}+\left[\chi_0\right]^{a\sigma_1,b\sigma_2}_{l\sigma'_1,m\sigma'_2}+\left[\chi_0\right]^{b\sigma_1,a\sigma_2}_{m\sigma'_1,l\sigma'_2}\right],\label{eq:BB_susc}\\
\left[\left\langle \mathcal{\hat{J}}_{\mathbf{q}} \mathcal{\hat{J}}_{\mathbf{-q}} \right\rangle\right]^{\sigma_1\sigma_2}_{\sigma'_1\sigma'_2}=\frac{1}{2}\sum_{(a,b),(lm)}Tr\left[
\left[\chi_0\right]^{a\sigma_1,b\sigma_2}_{m\sigma'_1,l\sigma'_2}+\left[\chi_0\right]^{b\sigma_1,a\sigma_2}_{l\sigma'_1,m\sigma'_2}-\left[\chi_0\right]^{a\sigma_1,b\sigma_2}_{l\sigma'_1,m\sigma'_2}-\left[\chi_0\right]^{b\sigma_1,a\sigma_2}_{m\sigma'_1,l\sigma'_2}\right].\label{eq:JJ_susc}
\end{gather}
For V-V channels: $(a,b),(l,m)\in \{(A,B), (B,C), (C,A) \}$, and for V-Sbip channels $(a,b),(l,m)\in \{(s,A), (s,B), (s,C) \}$. To calculate the BO \& LC susceptibilities in the charge and spin channels we contract the spin indices in Eqs.~\eqref{eq:BB_susc},~\eqref{eq:JJ_susc} as below,
\begin{gather}
    \chi^{c}_{\mathcal{BB}}(\mathbf{q})=\mathds{1}_{\sigma_1\sigma_2}\mathds{1}_{\sigma'_1\sigma'_2}\left[\left\langle \mathcal{\hat{B}}_{\mathbf{q}} \mathcal{\hat{B}}_{\mathbf{-q}} \right\rangle\right]^{\sigma_1\sigma_2}_{\sigma'_1\sigma'_2},\quad 
    \chi^{s}_{\mathcal{BB}}(\mathbf{q})=\frac{\Vec{\sigma}_{\sigma_1\sigma_2}\cdot\Vec{\sigma}_{\sigma'_1\sigma'_2}}{3}\left[\left\langle \mathcal{\hat{B}}_{\mathbf{q}} \mathcal{\hat{B}}_{\mathbf{-q}} \right\rangle\right]^{\sigma_1\sigma_2}_{\sigma'_1\sigma'_2},\\
    \chi^{c}_{\mathcal{JJ}}(\mathbf{q})=\mathds{1}_{\sigma_1\sigma_2}\mathds{1}_{\sigma'_1\sigma'_2}\left[\left\langle \mathcal{\hat{J}}_{\mathbf{q}} \mathcal{\hat{J}}_{\mathbf{-q}} \right\rangle\right]^{\sigma_1\sigma_2}_{\sigma'_1\sigma'_2},\quad 
    \chi^{s}_{\mathcal{JJ}}(\mathbf{q})=\frac{\Vec{\sigma}_{\sigma_1\sigma_2}\cdot\Vec{\sigma}_{\sigma'_1\sigma'_2}}{3}\left[\left\langle \mathcal{\hat{J}}_{\mathbf{q}} \mathcal{\hat{J}}_{\mathbf{-q}} \right\rangle\right]^{\sigma_1\sigma_2}_{\sigma'_1\sigma'_2}
\end{gather}
For the bare susceptibilities, we find that the charge and spin channels are always the same, $\chi^c_{\mathcal{BB}}(\mathbf{q})=\chi^s_{\mathcal{BB}}(\mathbf{q})$, $\chi^c_{\mathcal{JJ}}(\mathbf{q})=\chi^s_{\mathcal{JJ}}(\mathbf{q})$. The RPA-corrected BO \& LC susceptibilities are calculated by replacing the generalized bare-susceptibility $[\chi_0]$, with the generalized RPA-corrected susceptibility $\chi_{\text{RPA}}$, that is 
$\left[\chi_0\right]\rightarrow\left[\chi_{\text{RPA}}\right]$ in Eqs.~\eqref{eq:BB_susc},~\eqref{eq:JJ_susc}.

\subsection{Cooper vertex in band space projected into singlet \& triplet channels}
The Cooper vertex in orbital basis is written as,
\begin{gather}
H_{\text{Cooper}}=\frac{1}{2N}\sum_{\mathbf{k},\mathbf{k'},\{\Tilde{\mu}\}}\Tilde{V}(\mathbf{k},\mathbf{k'})^{\Tilde{\mu}_1\Tilde{\mu}_2}_{\Tilde{\mu}_3\Tilde{\mu}_4}c_{\mathbf{k},\Tilde{\mu}_1}^{\dagger}c_{\mathbf{-k},\Tilde{\mu}_3}^{\dagger}c_{-\mathbf{k'},\Tilde{\mu}_2}c_{\mathbf{k'},\Tilde{\mu}_4}\label{eq:Cooper_1}.
\end{gather}
We calculate $\Tilde{V}(\mathbf{k},\mathbf{k'})$ by  anti-symmetrizing the RPA-corrected vertex of Eq.~\eqref{eq:all_channels},
\begin{gather}
\Tilde{V}(\mathbf{k},\mathbf{k'})^{\Tilde{\mu}_1\Tilde{\mu}_2}_{\Tilde{\mu}_3\Tilde{\mu}_4}=\frac{1}{2}\left(V(\mathbf{k},\mathbf{k'})^{\Tilde{\mu}_1\Tilde{\mu}_2}_{\Tilde{\mu}_3\Tilde{\mu}_4}-V(\mathbf{k},\mathbf{-k'})^{\Tilde{\mu}_1\Tilde{\mu}_4}_{\Tilde{\mu}_3\Tilde{\mu}_2}\right)\label{eq:anti_symm_vertex}
\end{gather}
Accounting for the possibility of only intra-band pairing, this Cooper vertex in orbital space is transformed to band space as
\begin{gather}
H_{\text{Cooper}}=\frac{1}{2N}\sum_{\mathbf{k},\mathbf{k'},n_1,n_2\{\sigma\}}\Tilde{V}(n_1\mathbf{k},n_2\mathbf{k'})^{\sigma_1,\sigma_2}_{\sigma_3,\sigma_4}
c_{\mathbf{k}n_1 \sigma_1}^{\dagger}
c_{\mathbf{-k}n_1 \sigma_3}^{\dagger}
c_{-\mathbf{k'}n_2 \sigma_2}
c_{\mathbf{k'}n_2 \sigma_4},\label{eq:Cooper_projected}
\end{gather}
where $n$ is the band index. The band-basis and orbital-basis vertices are related by
\begin{equation}
  \Tilde{V}(n_1\mathbf{k},n_2\mathbf{k'})^{\sigma_1,\sigma_2}_{\sigma_3,\sigma_4}=\sum_{\{ \mu \}}M(\mathbf{k},\mathbf{k'})^{\Tilde{\mu}_1,\Tilde{\mu}_2,\Tilde{\mu}_3,\Tilde{\mu}_4}_{n_1,n_2}\Tilde{V}(\mathbf{k},\mathbf{k'})^{\Tilde{\mu}_1\Tilde{\mu}_2}_{\Tilde{\mu}_3\Tilde{\mu}_4},
\end{equation}
where this change of basis is carried out using
$M(\mathbf{k},\mathbf{k'})^{\Tilde{\mu}_1,\Tilde{\mu}_2,\Tilde{\mu}_3,\Tilde{\mu}_4}_{n_1,n_2}= \left[U^*(\mathbf{k}) \right]_{\mu_1 n_1} \left[U^*(\mathbf{-k}) \right]_{\mu_3 n_1} \left[U(\mathbf{-k'})\right]_{\mu_2 n_2} \left[U(\mathbf{k'}) \right]_{\mu_4 n_2}$. Next, in the band-basis, the Cooper vertex in the singlet and triplet channels are calculated using the projectors $\Gamma^0$ and $\{\Gamma^x, \Gamma^y, \Gamma^z \}$ respectively \cite{Romer_2019}. These are defined as: $\Gamma^{\eta}=\frac{i}{\sqrt{2}} \sigma^\eta\sigma^y$. The projectors act as follows, where $s_{\eta}=\{ -1,1,-1,1\}$
\begin{gather}  
\Tilde{V}_{\eta}(n_1\mathbf{k},n_2\mathbf{k'})=s_\eta\Gamma^{\eta}_{\sigma_3,\sigma_1}\Tilde{V}(n_1\mathbf{k},n_2\mathbf{k'})^{\sigma_1,\sigma_2}_{\sigma_3,\sigma_4} \Gamma^{\eta}_{\sigma_4,\sigma_2}\label{eq:sing_trip_projectors}.
\end{gather}
The singlet channel vertex $\Tilde{V}_{\eta=0}$, and the triplet channel vertices $\Tilde{V}_{\eta=x/y/z}$ are used in the linearized gap equation to obtain the leading and sub-leading superconducting instabilities in the singlet and triplet channels. The triplet channels are degenerate as spin-rotation symmetry is unbroken in the disordered state.

\section{Solving the gap equation}\label{sec:LGE_supp}
\subsection{Linearized gap equation}
To find the leading superconducting instability, we solve the linearized gap equation using the RPA-corrected Cooper vertex in the band-basis. We solve the linearized gap equation in both the singlet $(\eta=0)$ and triplet $(\eta=\{x,y,z\})$ channels. 
\begin{gather}
\lambda_{\eta}\Delta_{\eta}(n_1\mathbf{k})=-\frac{1}{(2\pi)^2}\sum_{n_2}\oint_{FS_{\mathbf{k'},n_2}}\frac{\Tilde{V}_{\eta}(n_1\mathbf{k},n_2\mathbf{k'}) \Delta_{\eta}(n_2\mathbf{k'}) d\mathbf{k'}}{\left|\nabla_{\mathbf{k'}}\xi_{\mathbf{k'}n_2}\right|}\label{eq:LGE_supp}
\end{gather} 
The gap function is defined as: $\Delta_{\eta}(n_1\mathbf{k})=\frac{1}{N}\sum_{\mathbf{k'},n_2}\Tilde{V}_{\eta}(n_1\mathbf{k},n_2\mathbf{k'})\left\langle \Psi_{\eta,n_2\mathbf{k'}} \right\rangle$, where $ \Psi_{\eta,n_2\mathbf{k'}}=\sum_{\sigma_1\sigma_2}\Gamma^{\eta}_{\sigma_2,\sigma_1}c_{\mathbf{-k},n_2,\sigma_2} c_{\mathbf{k},n_1,\sigma_2}$.
Each eigenvalue $\lambda_{\eta}>0$ gives a SC instability. The leading SC instability corresponds to the largest eigenvalue $\lambda_{\text{max}}$, with the gap structure across the Fermi surface encoded in the corresponding eigenvector $\Delta_{\text{max}}$. To solve the linearized gap equation we discretize the Fermi surface into $156$ points across the three sheets ($60+60+36$).

\subsection{Classifying the Superconducting Gaps}
\begin{figure}
    \centering
    \includegraphics[scale=0.75]{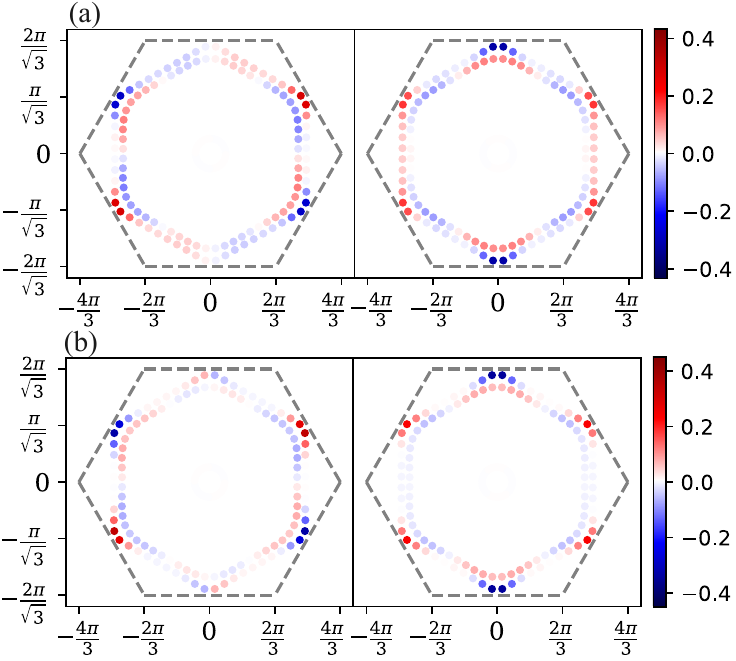}
    \caption{The $E_{2g}$ gap functions for fluctuating: (a) V-V cLC, (b) V-V sBO. The two gap functions have different weights on the lattice harmonics of $E_{2g}$ leading to a different gap structures, even though they belong to the same irrep.}
    \label{fig:E2_gaps}
\end{figure}

After obtaining the gap function $\Delta_{\eta}(n,\mathbf{k})$ across the Fermi surfaces from the linearized gap equation in Eq.~\eqref{eq:LGE_supp}, we classify them based on the lattice harmonics of the $D_{6h}$ point group \cite{Hoelbaek_thesis}. Since we only focus on the $k_z=0$ states, we list the lattice harmonics of $D_{6h}$ having a non-zero value on the $k_z=0$ plane in Table \ref{tab:lattice_harmonics}. The lattice harmonics, corresponding to vector $\mathbf{r}$, are generated by projecting the basis functions $e^{i\mathbf{k}\cdot\mathbf{r}}$ onto the irreps of the point group,
\begin{gather}
    %b_{\mathbf{r}}^{\Gamma}(\mathbf{k})=\frac{1}{N_G}\sum_{g \in G}\chi^{\Gamma}(g) e^{i\mathbf{k\cdot (D(g)\mathbf{r})}}\nonumber\\
    b_{\mathbf{r}}^{\Gamma}(\mathbf{k})_{\mu\nu}=\frac{1}{N_G}\sum_{g \in G}D^{\Gamma}(g)^{*}_{\mu\nu} e^{i\mathbf{k\cdot (D(g)\mathbf{r})}}\label{eq:lattice_harmonic}.
\end{gather}
For the 1D irreps of $D_{6h}$: $\{A_{1g},A_{2g},B_{1u},B_{2u}\}$, we use the characters of the point group, i.e.  $D^{\Gamma}(g)^*=\chi^{\Gamma}(g)$ as the irreps for the projection. For the 2D irreps: $\{E_{1u},E_{2g}\}$, we use generators of the 2D representation of $D_{6h}$ for the projection. For the 2D irreps, we take the $(00)$, and $(01)$ components to construct the lattice harmonics. The lowest non-zero lattice harmonics of each irrep are usually called s-, p-, d-, ... wave states. The presence of higher-harmonics in the gap function may lead to additional zeros in the gap. For example, in Fig.~\ref{fig:E2_gaps} we show two different gap functions belonging to the $E_{2g}$ irrep that have different weights on the higher-harmonics.

\begin{table}[h!]
    \centering
    \begin{tabular}{|c|c|c|c|c|}%{|p{0.8cm}|p{2cm}|p{1.2cm}|p{4cm}|p{9.5cm}|}
        \hline
         Irrep.  &  Lowest order  & Gap function of lowest harmonic in $\mathbf{k}$-space\\
           &  Polynomial functions  & \\
         \hline
         \hline
         $A_{1g}$   & s & 1 \\
         \hline
         $A_{2g}$  & $i_{xy(x^2-3y^2)(3x^2-y^2)}$ & $-\frac{16}{3}\sin\!\left(\frac{k_x}{2}\right)\sin\!\left(\frac{\sqrt{3}k_y}{2}\right)\sin\!\left(\frac{k_x-\sqrt{3}k_y}{4}\right)$ $\sin\!\left(\frac{3k_x-\sqrt{3}k_y}{4}\right)\sin\!\left(\frac{k_x+\sqrt{3}k_y}{4}\right)\sin\!\left(\frac{3k_x+\sqrt{3}k_y}{4}\right)$\\
         \hline
         $B_{1u}$  & $f_{y(3x^2-y^2)}$ & $\frac{2}{3}\left(\cos{\left(\frac{k_x}{2}\right)} - \cos{\left(\frac{\sqrt{3}k_y }{2}\right)}\right) \sin{\left(\frac{k_x}{2}\right)}$\\
         \hline
         $B_{2u}$ & $f_{x(x^2-3y^2)}$  & $\frac{2}{3}\left(\cos\left(\frac{3k_x}{2}\right)-\cos\left(\frac{\sqrt{3}k_y}{2}\right)\right)\sin\left(\frac{\sqrt{3}k_y}{2}\right)$\\
         \hline
         $E_{1u}$  & $p_x,p_y$ & $\frac{1}{3}\left(2\cos\left(\frac{k_x}{2}\right)+\cos\left(\frac{\sqrt{3}k_y}{2}\right)\right)\sin\left(\frac{k_x}{2}\right), \frac{1}{\sqrt{3}}\cos\left(\frac{k_x}{2}\right)\sin\left(\frac{\sqrt{3}k_y}{2}\right) $\\
         \hline
         $E_{2g}$   & $d_{x^2-y^2},d_{xy}$ & $ \frac{1}{3}\left(\cos(k_x)-\cos\left(\frac{k_x}{2}\right)\cos\left(\frac{\sqrt{3}k_y}{2}\right)\right), -\frac{1}{\sqrt{3}}\sin\left(\frac{k_x}{2}\right)\sin\left(\frac{\sqrt{3}k_y}{2}\right)$\\
        \hline
    \end{tabular}
    \caption{Lowest lattice harmonics for $D_{6h}$ that are non-zero for $k_z=0$. $A_{1g}$, $A_{2g}$, $E_{2g}$ are singlet gaps, and $B_{1u}$, $B_{2u}$, $E_{1u}$ are triplet gap functions.}
    \label{tab:lattice_harmonics}
\end{table}

\section{Time-reversal parity of fluctuating mode and singlet Cooper pairing}\label{sec:T_parity_supp}
\subsection{A simple proof}
Previous studies \cite{Scheurer2016,Palle_thesis} have shown that fluctuations of a time-reversal even (odd) mode mediates conventional (unconventional) SC. In this section, we give a simple proof of this fact for the benefit of the reader.

We consider a fluctuating order parameter $\mathcal{\hat{O}}_\mathbf{q}^i$ (with no spin-orbit coupling) having a definite parity  $P_T$ ($\pm1$) under time reversal, $\mathcal{T}\mathcal{\hat{O}}_\mathbf{q}^i \mathcal{T}^{-1}=P_T \mathcal{\hat{O}}_\mathbf{-q}^i$. The order parameter is defined for $i\in \{0,x,y,z \}$ as follows
\begin{gather}
\mathcal{\hat{O}}_\mathbf{q}^{i}=\sum_{\mathbf{k}}\mathcal{O}_{\mu_1 \mu_2}(\mathbf{k},\mathbf{k+q})\sigma^i_{\sigma_1 \sigma_2} c_{\mathbf{k}\mu_1\sigma_1}^{\dagger}c_{\mathbf{k+q}\mu_2\sigma_2}
 .   \label{eq:defn_of_O_q^i}
\end{gather}
Under time-reversal,
$\mathcal{T} c_{\mathbf{k}\mu_1\sigma_1} \mathcal{T}^{-1}=i(\sigma^y)_{\sigma_1\sigma_2}c_{-\mathbf{k}\mu_1\sigma_2}$, and $\mathcal{T} c^{\dagger}_{\mathbf{k}\mu_1 \sigma_1} \mathcal{T}^{-1}=i(\sigma^y)_{\sigma_1 \sigma_2}c^{\dagger}_{-\mathbf{k}\mu_1 \sigma_2}$. This implies 
\begin{gather}
    \mathcal{T} \mathcal{\hat{O}}_\mathbf{q}^{i} \mathcal{T}^{-1}
    =\sum_{\mathbf{k}} c_{\mathbf{k}\mu_1s_1'}^{\dagger}\mathcal{O}^{*}_{\mu_1 \mu_2}(\mathbf{-k},\mathbf{-k+q})\sigma^y_{s'_1 \sigma_1}\sigma^{*i}_{\sigma_1 \sigma_2}  \sigma^y_{\sigma_2 s'_{2} }c_{\mathbf{k-
    q}\mu_2 s_2'}.\label{eq:time_rev_O_q^i}
\end{gather}
As $\mathcal{\hat{O}}_\mathbf{q}^{i}$ has a definite parity under time-reversal $P_T$, we can find using Eqs.~\eqref{eq:defn_of_O_q^i},~\eqref{eq:time_rev_O_q^i} that
\begin{gather}
    P_T \mathcal{O}_{\mu_1 \mu_2}(\mathbf{k},\mathbf{k-q})\sigma^{i}_{s'_1 s'_2}  =\mathcal{O}^{*}_{\mu_1 \mu_2}(\mathbf{-k},\mathbf{-k+q})\sigma^y_{s'_1 \sigma_1}\sigma^{*i}_{\sigma_1 \sigma_2}  \sigma^y_{\sigma_2 s'_2} \label{eq:parity_o_q^i}.
\end{gather}
Since $\mathcal{\hat{O}}^i_{\mathbf{q}}$ is the fluctuating mode the vertex can be written as $W=-\frac{g^2}{2N            }\sum_{\mathbf{q}}\mathcal{\hat{O}}_\mathbf{q}^{i} \mathcal{\hat{O}}_\mathbf{-q}^{i}$. In the Cooper channel of the orbital-basis it can be written as
\begin{gather}
V(\mathbf{k},\mathbf{k'})^{\Tilde{\mu}_1 \Tilde{\mu}_2}_{\Tilde{\mu}_3 \Tilde{\mu}_4}=-\frac{g^2}{2}\mathcal{O}_{\mu_1 \mu_4} (\mathbf{k},\mathbf{k'})\mathcal{O}_{\mu_3\mu_2}(\mathbf{-k},\mathbf{-k'}) \sigma^i_{\sigma_1 \sigma_4} \sigma^i_{\sigma_3 \sigma_2}\label{eq:cooper_PT}.
\end{gather}
Anti-symmetrizing the vertex in Eq.~\eqref{eq:cooper_PT}, and changing it into the band-basis and using Eq.~\eqref{eq:parity_o_q^i} we find
\begin{gather}
\Tilde{V}(n_1\mathbf{k},n_2\mathbf{k'})^{\sigma_1\sigma_2}_{\sigma_3\sigma_4}
=-\frac{g^2 P_T}{4}\left( \left|\mathcal{O}_{n_1 n_2} (\mathbf{k},\mathbf{k'})\right|^2 \sigma^i_{\sigma_1 \sigma_4} \left[\sigma^y\sigma^{*i}\sigma^y\right]_{\sigma_3 \sigma_2}-\left|\mathcal{O}_{n_1 n_2} (\mathbf{k},\mathbf{-k'})\right|^2 \sigma^i_{\sigma_1 \sigma_2} \left[\sigma^y\sigma^{*i}\sigma^y\right]_{\sigma_3 \sigma_4}\right)
\end{gather}
where $\mathcal{O}_{n_1 n_2} (\mathbf{k},\mathbf{k'})=\sum_{\mu_1 \mu_2}\left[U^*(\mathbf{k})\right]_{\mu_1 n_1}\mathcal{O}_{\mu_1 \mu_2} (\mathbf{k},\mathbf{k'}) \left[U(\mathbf{k})\right]_{\mu_2 n_2}$, and $n_1$ and $n_2$ are band indices. The Cooper vertex in the singlet channel can be written as
\begin{gather}
\Tilde{V}_{\eta=\text{0}}(n_1\mathbf{k},n_2\mathbf{k'})=-\frac{g^2P_T}{2}\left( \left|\mathcal{O}_{n_1n_2} (\mathbf{k},\mathbf{k'})\right|^2 +\left|\mathcal{O}_{n_1n_2} (\mathbf{k},\mathbf{-k'})\right|^2 \right),\label{eq:singlet_parity]}
\end{gather}
and in the triplet channel
\begin{gather}
\Tilde{V}_{\eta=\text{x/y/z}}(n_1\mathbf{k},n_2\mathbf{k'})=-\frac{g^2P_T}{2}\left( \left|\mathcal{O}_{n_1n_2} (\mathbf{k},\mathbf{k'})\right|^2 -\left|\mathcal{O}_{n_1n_2} (\mathbf{k},\mathbf{-k'})\right|^2 \right).
\label{eq:triplet_parity}
\end{gather}
Eq.~\eqref{eq:singlet_parity]} is written using a compact notation in the main text. The singlet channel Cooper vertex, Eq.~\eqref{eq:singlet_parity]}, implies that a time-reversal even ($P_T=+1$) fluctuating mode is attractive for all pairs of $(\mathbf{k},\mathbf{k'})$ on the Fermi surface, and a time-reversal odd ($P_T=-1$) mode is repulsive for all pairs of $(\mathbf{k},\mathbf{k'})$ on the Fermi surface. However, from Eq.~\eqref{eq:triplet_parity}, we see that a same claim cannot be made for the triplet channel. From the linearized gap equation in~\eqref{eq:LGE_supp}, we see that in the singlet channel a repulsive interaction, $\Tilde{V}(n_1\mathbf{k},n_2\mathbf{k'})>0$, necessitates a sign change in the gap structure leading to an unconventional pairing symmetry. However, for an attractive channel $\Tilde{V}(n_1\mathbf{k},n_2\mathbf{k'})<0$, $\Delta(\mathbf{k})$ and $\Delta(\mathbf{k'})$ have the same sign, thus leading to conventional s-wave pairing
.
\subsection{Vertices with competing modes having different time-reversal parity}
A vertex can be made up of both the charge and spin channels of a fluctuating mode $\mathcal{\hat{O}_{\mathbf{q}}}$ as below,
\begin{gather}
H_{\text{int}}=-\frac{g^2}{2} \left( \left(1-\alpha\right) \mathcal{\hat{O}}_\mathbf{q}^{c} \mathcal{\hat{O}}_\mathbf{-q}^{c}
+\alpha\mathcal{\hat{O}}^{s}_\mathbf{q} \cdot\mathcal{\hat{O}}^{s}_\mathbf{-q}
\right)\label{eq:mixed_vertex_gen}.
\end{gather}
Here we have used the notation of Eq.~\eqref{eq:final_interaction_ham_split}. We assume $P_T$ is the time-reversal parity of the mode in the charge channel $\mathcal{\hat{O}}_\mathbf{q}^{c}$. The mode in the spin channel $\mathcal{\hat{O}}^{s}_\mathbf{q}$ has the opposite $P_T$. Projecting this mixed charge-spin sector vertex in Eq.~\eqref{eq:mixed_vertex_gen} in the singlet-Cooper channel we find
\begin{gather}
\Tilde{V}_{\eta=\text{0}}(n_1\mathbf{k},n_2\mathbf{k'})=-\frac{g^2P_T}{2}\left(1-4\alpha\right)\left( \left|\mathcal{O}_{n_1n_2} (\mathbf{k},\mathbf{k'})\right|^2 +\left|\mathcal{O}_{n_1n_2} (\mathbf{k},\mathbf{-k'})\right|^2 \right)
\label{eq:mixed_vertex_gen_singlet}  
\end{gather}
Therefore, we expect a transition in the pairing symmetry in the singlet channel to happen at $\alpha=\frac{1}{4}$. In Fig. 3 (a) of the main text, we see transitions in the pairing symmetry happening close to $\alpha=\frac{1}{4}$, which is consistent with this result. The slight shift in the phase boundary to $\alpha>\frac{1}{4}$ can be explained from the RPA-correction to the Cooper vertex as described below.
\subsection{Adding RPA-corrections}
In the above, sections we showed that fluctuations of $P_T=+1(-1)$ modes mediate conventional (unconventional) singlet pairing. We proved this at the level of the bare vertices. We now show the effect of adding RPA-corrections to this result. As in the previous calculation, we assume that the vertex is made up of both the charge and spin channels similar to Eq.~\eqref{eq:mixed_vertex_gen}. For simplicity, we assume that the orbital-sector of the vertex is a constant having no momentum dependence. With this assumption, we can drop the orbital indices in the vertex. Therefore, the bare vertex can be written in with only the spin sector as
\begin{gather}
\left[V_{\text{bare}}\right]^{\sigma_1\sigma_2}_{\sigma_3\sigma_4}=-W \left( \left(1-\alpha\right) \mathds{1}_{\sigma_1\sigma_4}\mathds{1}_{\sigma_3\sigma_2}
+\alpha \Vec{\boldsymbol{\sigma}}_{\sigma_1\sigma_4}\cdot \Vec{\boldsymbol{\sigma}}_{\sigma_3\sigma_2}
\right),
\end{gather}
where $W=\frac{g^2\mathcal{O}}{2}$, and $\mathds{1}=\sigma^0$. Upon including RPA-corrections, the vertex becomes
\begin{gather}
[V_{\text{RPA-corrected}}]^{\sigma_1\sigma_2}_{\sigma_3\sigma_4}=-W\left(\frac{(1-\alpha)}{1-2W\chi (1-\alpha)}\mathds{1}_{\sigma_1\sigma_4}\mathds{1}_{\sigma_3\sigma_2}+\frac{\alpha}{1-2W\chi \alpha} \Vec{\boldsymbol{\sigma}}_{\sigma_1\sigma_4}\cdot \Vec{\boldsymbol{\sigma}}_{\sigma_3\sigma_2}\right).\label{eq:RPA_corrected_mixed_vertex}
\end{gather}
Further, similar to Eq.~\eqref{eq:mixed_vertex_gen_singlet} we can write down this vertex in Eq.~\eqref{eq:RPA_corrected_mixed_vertex} projected into the singlet channel as:
\begin{gather}
    \Tilde{V}_0=-\frac{g^2 P_T}{2}\left(\frac{1-\alpha}{1-2W \chi(1-\alpha)}-\frac{3\alpha}{1-2W\chi\alpha}\right) 2|\mathcal{O}|^2\label{eq:RPA_vertex_singlet}.
\end{gather}
From Eq.~\eqref{eq:mixed_vertex_gen_singlet}, we found that a transition in the pairing instability happens at $\alpha=\frac{1}{4}$ for the bare vertices. However, with the heuristic form of the RPA-corrected vertex in from Eq.~\eqref{eq:RPA_vertex_singlet}, we can determine that the transition now occurs for $\alpha>\frac{1}{4}$, which is consistent with our findings in Figs. 3(a,b) of the main-text.

We can also write down a similar orbital-independent form of the charge and spin susceptibilities,
\begin{gather}
\left[\chi_{c,\text{RPA}}\right]=\delta_{\sigma_1\sigma_2}\delta_{\sigma_3\sigma_4}\left[\chi_{\alpha,\text{RPA}}\right]^{\sigma_1\sigma_2}_{\sigma_3\sigma_4}=\frac{2\chi}{1-2W \chi (1-\alpha)},\quad
\left[\chi_{s,\text{RPA}}\right]=\frac{\Vec{\sigma}_{\sigma_1\sigma_2}\cdot\Vec{\sigma}_{\sigma_3\sigma_4}}{3}\left[\chi_{\alpha,\text{RPA}}\right]^{\sigma_1\sigma_2}_{\sigma_3\sigma_4}=\frac{2\chi}{ 1-2W\chi \alpha}\label{eq:orb_ind_susc}.
\end{gather}
Eq.~\eqref{eq:orb_ind_susc} explains the equality for the spin and charge susceptibilities at $\alpha=\frac{1}{2}$ as we saw in Fig. 2 of the main text. This is caused by the inclusion of only the RPA-bubble diagrams.

\section{Visualizing the singlet and triplet vertices}\label{sec:visualizing_vertices}
\begin{figure}
\centering\includegraphics[scale=0.55]{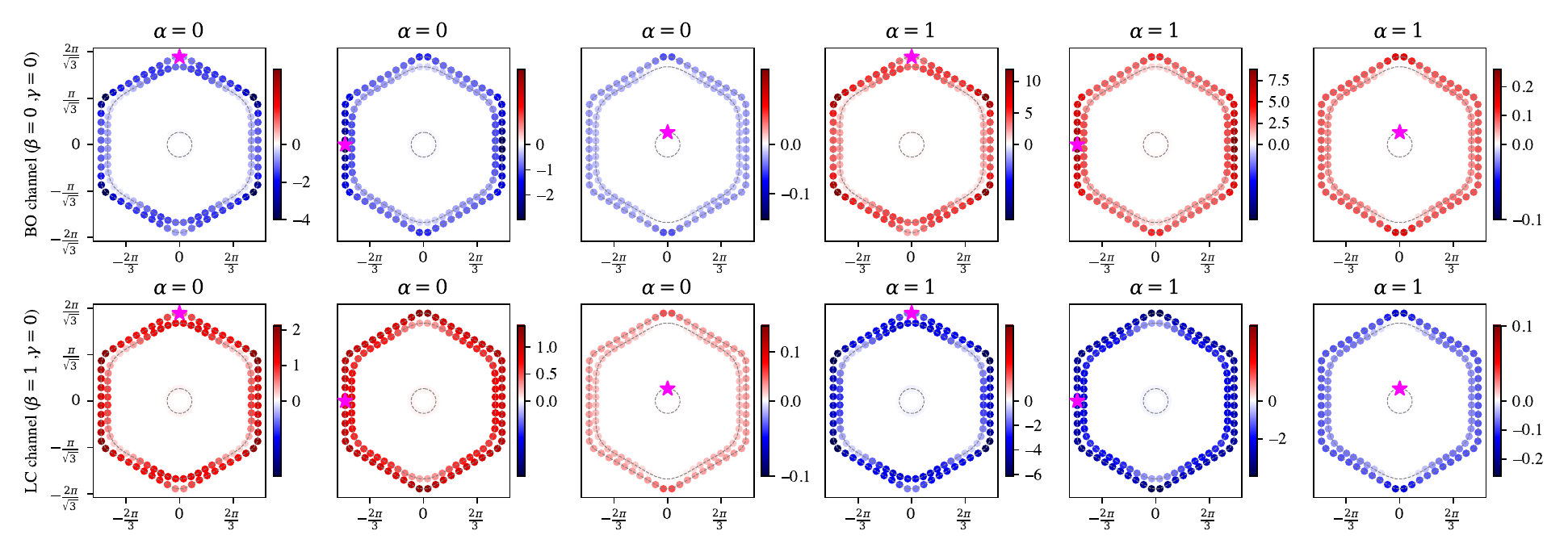}
    \caption{The RPA-corrected singlet Cooper vertex $\Tilde{V}_{\eta=0}(\mathbf{k^*},\mathbf{k})$ plotted with $\mathbf{k^*}$ fixed at the momenta shown with pink star. The vertices are shown for a the fluctuating V-V BOs (top row), and V-V LCs (bottom row) in the $\alpha=0$ (charge) and $\alpha=1$ (spin) sectors. The figure is shown for $\mu=3.88$ and $\xi=4$.}
    \label{fig:singlet_VV}
\end{figure}

\begin{figure}
\centering\includegraphics[scale=0.55]{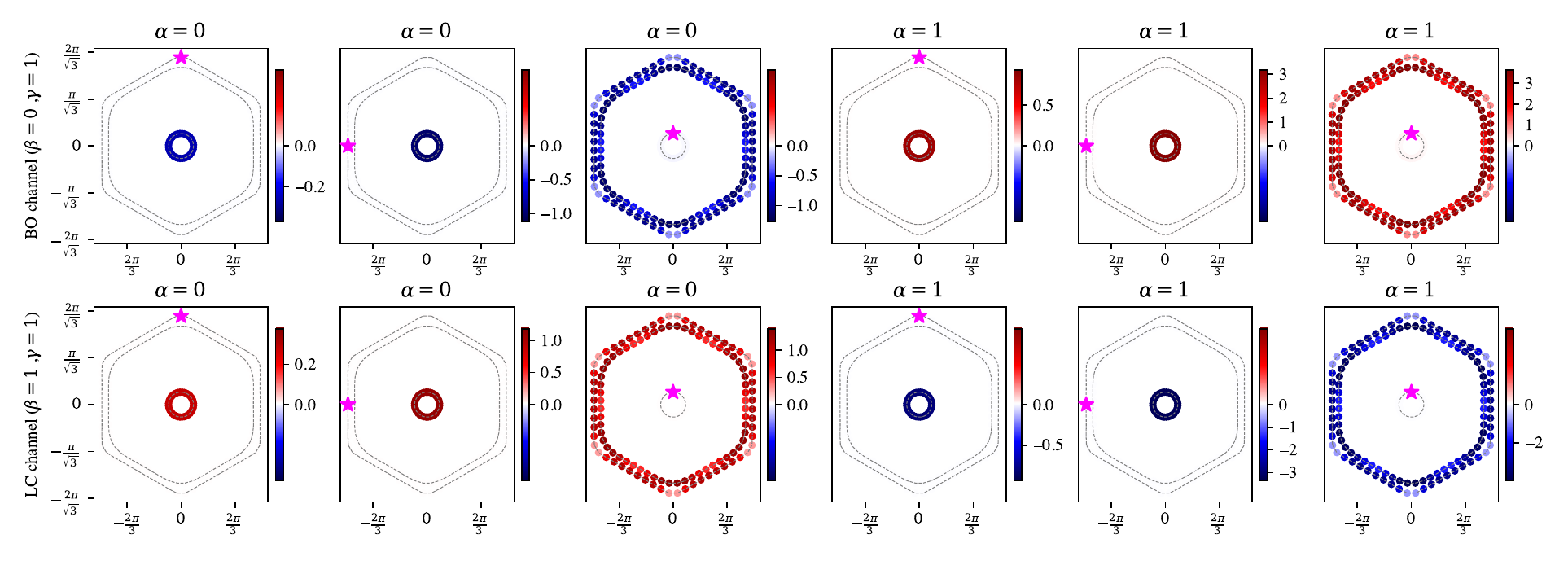}
    \caption{The RPA-corrected singlet Cooper vertex $\Tilde{V}_{\eta=0}(\mathbf{k^*},\mathbf{k})$ plotted with $\mathbf{k^*}$ fixed at the momenta shown with pink star. The vertices are shown for a the fluctuating V-Sb-ip BOs (top row), and V-Sb-ip LCs (bottom row) in the $\alpha=0$ (charge) and $\alpha=1$ (spin) sectors. The figure is shown for $\mu=3.88$ and $\xi=4$.}
    \label{fig:singlet_VSbip}
\end{figure}

\begin{figure}
\centering\includegraphics[scale=0.55]{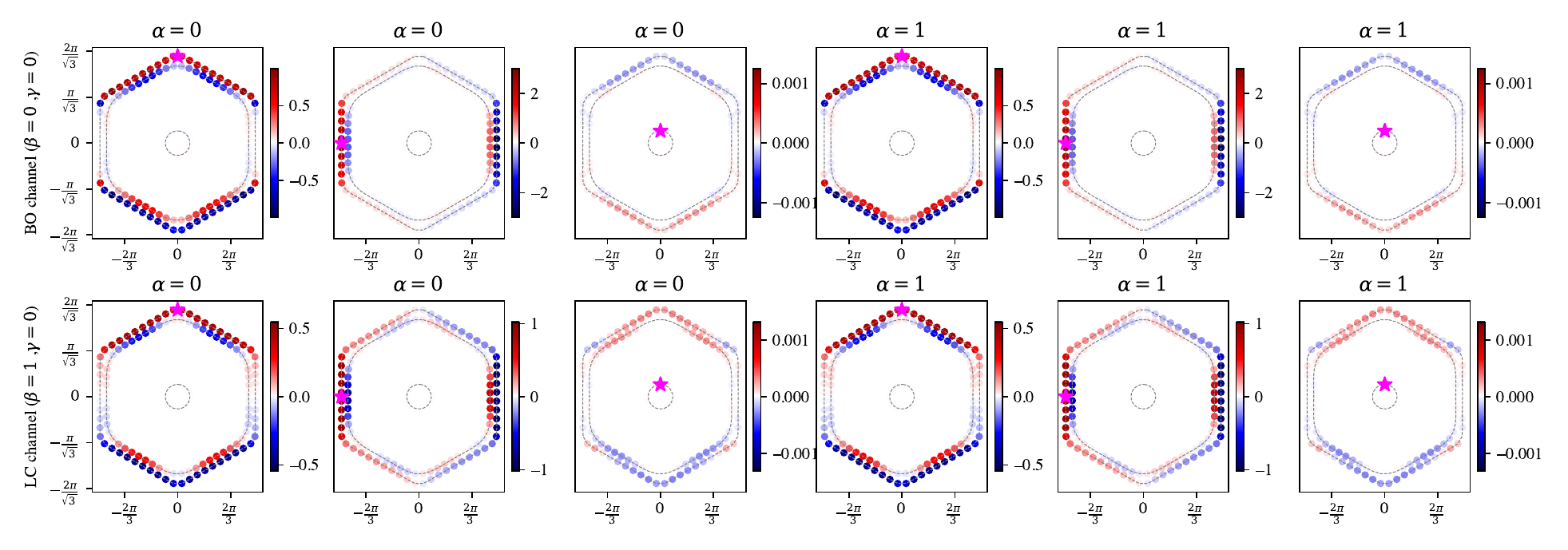}
    \caption{The RPA-corrected triplet Cooper vertex $\Tilde{V}_{\eta=x/y/z}(\mathbf{k^*},\mathbf{k})$ plotted with $\mathbf{k^*}$ fixed at the momenta shown with pink star. The vertices are shown for a the fluctuating V-V BOs (top row), and V-V LCs (bottom row) in the $\alpha=0$ (charge) and $\alpha=1$ (spin) sectors. The figure is shown for $\mu=3.88$ and $\xi=4$.}
    \label{fig:triplet_VV}
\end{figure}

\begin{figure}
\centering\includegraphics[scale=0.55]{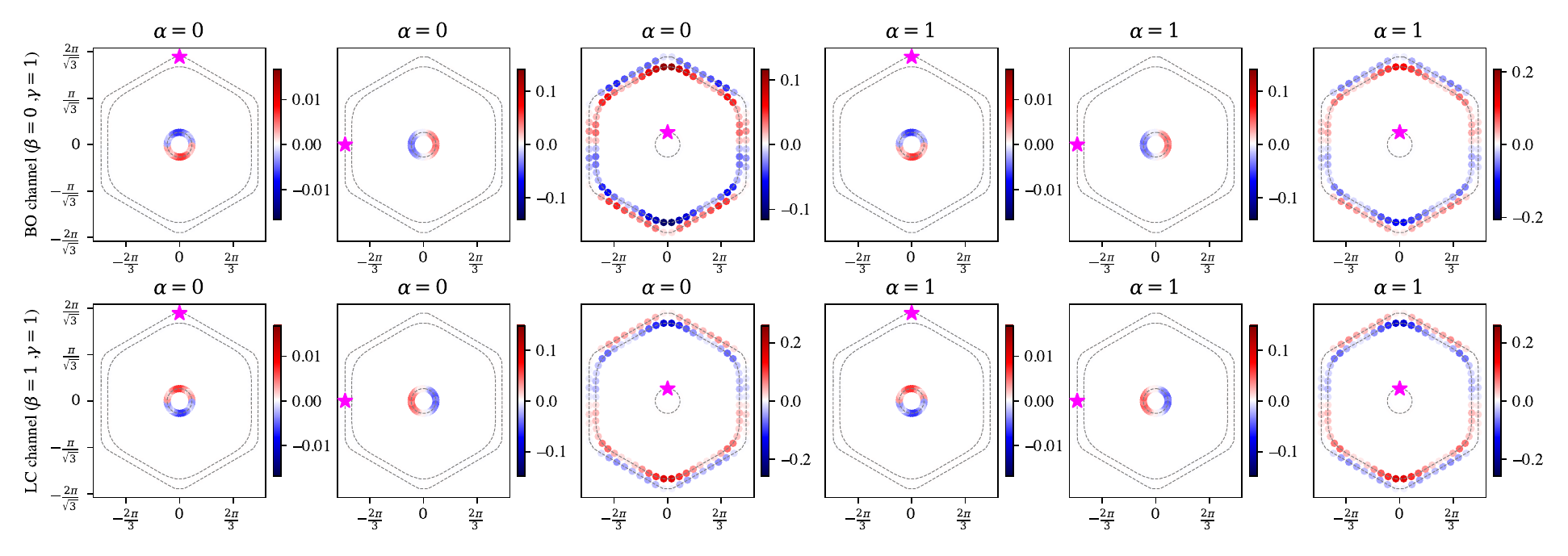}
    \caption{The RPA-corrected triplet Cooper vertex $\Tilde{V}_{\eta=x/y/z}(\mathbf{k^*},\mathbf{k})$ plotted with $\mathbf{k^*}$ fixed at the momenta shown with pink star. The vertices are shown for a the fluctuating V-Sb-ip BOs (top row), and V-Sb-ip LCs (bottom row) in the $\alpha=0$ (charge) and $\alpha=1$ (spin) sectors. The figure is shown for $\mu=3.88$ and $\xi=4$.}
    \label{fig:triplet_VSbip}
\end{figure}

In Figs.~\ref{fig:singlet_VV},~\ref{fig:singlet_VSbip}, we show the RPA-corrected singlet Cooper vertices for fluctuating V-V and V-Sbip modes respectively. To plot the singlet vertex, $\Tilde{V}_{\eta=0}(\mathbf{k^*},\mathbf{k})$, we fix $\mathbf{k^*}$ to points along a high-symmetry direction in the Brillouin zone. We see from Figs.~\ref{fig:singlet_VV},~\ref{fig:singlet_VSbip}, that the time-reversal even modes: cBO and sLC are attractive for all $(\mathbf{k}^*,\mathbf{k})$ and therefore lead to conventional s-wave pairing as we found from Fig. 3 (a) of the main text. Additionally, we see that time-reversal odd modes: sBO and cLC are repulsive for all $(\mathbf{k}^*,\mathbf{k})$, leading to unconventional chiral $d+id$ or $s_{+-}$ pairing. Similarly, in Figs.~\ref{fig:triplet_VV},~\ref{fig:triplet_VSbip}, we show the RPA-corrected triplet Cooper vertices, $\Tilde{V}_{\eta=x/y/z}(\mathbf{k^*},\mathbf{k})$. The three triplet channels: $\eta=x/y/z$ are degenerate and all lead to the same pattern shown in Figs.~\ref{fig:triplet_VV},~\ref{fig:triplet_VSbip}.

\section{Near the M-type VHS}\label{sec:5.02}
\subsection{Susceptibility}
\begin{figure}
    \centering
    \includegraphics[scale=0.85]{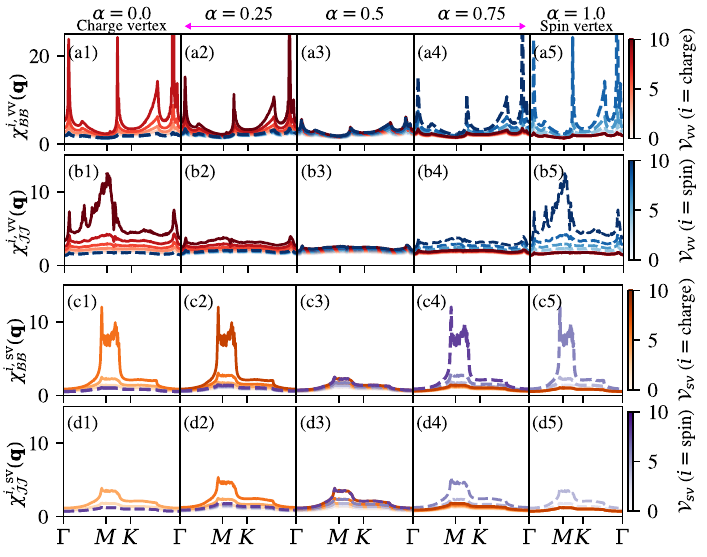}
    \caption{The RPA corrected susceptibilities (at $\mu=5.02$) for: bond-orders $\chi_{\mathcal{BB}}(\mathbf{q})$ ((a) between V-V, (c) between V-Sbip), and loop-currents $\chi_{\mathcal{JJ}}(\mathbf{q})$ ((b) between V-V, (d) between V-Sbip). Columns 1-5 show the susceptibilities in the charge and spin channels with increasing interaction strengths (shades of colors) as the relative strength of the charge and spin vertex is tuned by $\alpha$. Only the $\chi$’s in the disordered state are shown, some larger interaction strengths (or darker color shades) are omitted as those channels would condense.}
    \label{fig:susc_5.02}
\end{figure}

In the main text, we showed the leading SC instability when the chemical potential is close to the P-type VHS ($\mu=3.88$). In this section we present the results for $\mu=5.02$, which is closer to the M-type VHS (see Fig.~\ref{fig:1_supp} (c)). For $\mu=5.02$, the V-V LC susceptibility, and V-Sbip BO \& LC susceptibilities, as shown in Fig.~\ref{fig:susc_5.02}, are strongly enhanced near the all the three $M_{c}$-points (where $c\in \{A,B,C\}$). However, the V-V CBO susceptibility shows a tendency to condense at a different wave vector. 
\subsection{SC phases}
\begin{figure}
    \centering
\includegraphics[scale=0.75]{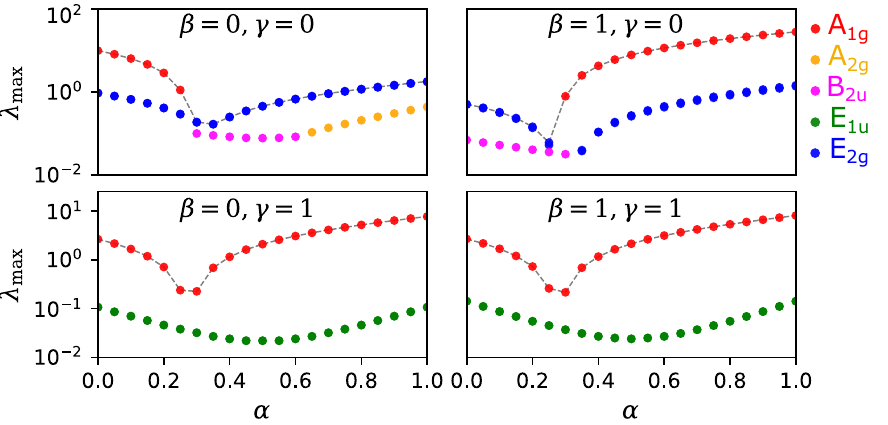}
    \caption{The SC instabilities for $\mu=5.02$ (near the M-type VHS) are shown as a function of $\alpha$, which tunes between the charge and spin channels. The panels correspond to fluctuations of: V-V BO ($\beta=\gamma=0$), V-V LC ($\beta=1$, $\gamma=0$), V-Sbip BO ($\beta=0$, $\gamma=1$), V-Sbip LC ($\beta=\gamma=1$). We find the same leading pairing instabilities as in Fig. 3(c) of the main text. All results in this figure are obtained for $\xi=4$.}
    \label{fig:SC_5.02}
\end{figure}

We find that moving the chemical potential to $\mu=5.02$ yields a phase diagram qualitatively similar to that at $\mu=3.88$ [Fig. 3 (a) of the main text], where singlet pairing predominates. In Fig.~\ref{fig:SC_5.02}, we show some paths across the SC phase diagram. As discussed earlier, the pairing symmetry of the singlet is primarily dictated by the TR parity of the strongest fluctuations. Therefore, we would expect the pairing symmetry of the singlet channel to be unchanged for $\mu=5.02$. This is consistent with our findings shown in Fig.~\ref{fig:SC_5.02}.

